\documentclass[11pt,]{article}
\usepackage{authblk}
\usepackage{fullpage}
\usepackage{amssymb,amsmath}
\usepackage[utf8x]{inputenc}
\usepackage[T1]{fontenc}
\usepackage{siunitx}
\usepackage[version=3]{mhchem}
\usepackage{xcolor}
\usepackage{natbib}
 \usepackage{mathtools}
 \usepackage{booktabs}
 \usepackage{multirow}

\usepackage[a4paper, total={6in, 8in}]{geometry}

\newcommand*{\fref}[1]{{figure~\ref{#1}}} % per fare riferimento a figure
\usepackage{setspace}
\usepackage{secdot} % Dot after section number
\usepackage{siunitx}

\usepackage[bookmarks=false,hyperfootnotes=false]{hyperref}
\hypersetup{
			colorlinks=true,
			linkcolor=blue,
			anchorcolor=black,
			citecolor=blue,
			urlcolor=blue,
}

\usepackage{graphicx}
\graphicspath{{}}
\makeatletter
\def\ScaleWidthIfNeeded{%
 \ifdim\Gin@nat@width>\linewidth
    \linewidth
  \else
    \Gin@nat@width
  \fi
}
\def\ScaleHeightIfNeeded{%
  \ifdim\Gin@nat@height>0.9\textheight
    0.9\textheight
  \else
    \Gin@nat@width
  \fi
}
\makeatother
\setkeys{Gin}{width=\ScaleWidthIfNeeded,height=\ScaleHeightIfNeeded,keepaspectratio}%

\title{Effect of free-stream turbulence on a moderate adverse pressure gradient turbulent boundary layer developing over an airfoil}
\author[$\ast$ 2]{Tomek Jaroslawski}
\author[1]{Francesco Scarano}

\affil[1]{DIMEAS Politecnico di Torino.}
\affil[2]{Center for Turbulence Research, Stanford University.}

\date{\today}

\begin{document}

\maketitle

%\newpage
\section*{Abstract}\label{sec:abstract}
Turbulent boundary layers (TBLs) subjected to adverse pressure gradients (APGs) are highly relevant to industrial aerodynamic applications, yet the effect of freestream turbulence (FST) on TBLs developing under moderate APGs remains insufficiently understood. To address this, we conducted wind tunnel experiments to investigate the effects of FST on a developing TBL over a NACA 0015 airfoil. Varying the angle of attack (2 and 4$^\circ$) adjusted the pressure gradient, and hotwire anemometry measured boundary layer properties at different chordwise positions ($x/c$ = 0.400--0.625, with $\beta = \delta^*/\tau_0 dP/dx$ = 0.2--1.5). The FST level was increased using static grids, resulting in levels ranging from 0.15 to 6$\%$. The chord-based Reynolds number was kept constant at around 250,000 for all configurations. The results show that increasing FST systematically modifies the mean-flow development and turbulence structure of the APG boundary layer. Higher FST levels reduce the shape factor and partially suppress the APG-induced wake in the mean velocity profile, while increasing the skin-friction coefficient towards values closer to canonical ZPG behaviour. The streamwise velocity variance is amplified in both the inner and outer regions, and spectral analysis shows that this increase is associated with energetic large-scale motions introduced by the freestream turbulence, with characteristic wavelengths of order $\lambda_x/\delta \approx 13$. These motions penetrate into the boundary layer and contribute to the near-wall variance, with a stronger effect observed as the adverse pressure gradient increases. 
The findings demonstrate that FST is a governing parameter in spatially developing APG TBLs over airfoils, since its influence appears to be amplified by the adverse pressure gradient, and must therefore be accounted for when interpreting mean-flow development, turbulence statistics and scale interactions in realistic aerodynamic environments.

\vskip10pt

$\ast$\href{mailto:me@example.com}{tomek@stanford.edu}

\newpage

\section{Introduction}\label{sec:introduction}
Turbulent boundary layers (TBLs) subjected to adverse pressure gradients (APGs) are central to many aerodynamic applications, where the local deceleration of the flow promotes wake growth, reduced skin friction and, eventually, separation. In practical configurations, however, APG TBLs rarely develop under clean free-stream conditions. Airfoils operating in wind-turbine wakes, turbomachinery passages, fan flows, atmospheric inflow and urban environments are often exposed to elevated levels of free-stream turbulence (FST) \citep{gambuzza2021effects,volino1997new,jaroslawski2023boundary}. Similar conditions may also arise in wind-tunnel experiments, where residual or grid-generated FST can contribute to differences between measurements and numerical simulations performed under nominally clean inflow. Understanding how FST modifies APG TBLs is therefore relevant both for flow-physics modelling and for the prediction of aerodynamic performance and separation over aerodynamic bodies.

The effect of FST on zero-pressure-gradient (ZPG) TBLs has been extensively investigated. Here, the streamwise turbulence intensity is denoted by
\begin{equation}
T_u = u'_\infty/U_\infty,
\end{equation}
where $u'_\infty$ is the root-mean-square of the streamwise velocity fluctuations in the free stream and $U_\infty$ is the free-stream velocity. Early experiments by \citet{hancock1983effect,hancock1989turbulence} considered FST levels up to $T_u=5.8\%$ and integral length-scale ratios $L_u/\delta=0.67$--$2.23$, where $L_u$ is the characteristic size of the energy-containing free-stream motions and $\delta$ is the boundary-layer thickness. They showed that both $T_u$ and $L_u/\delta$ influence the wall shear stress and the wake region, and proposed combined FST parameters to correlate these effects. However, some measurements were performed relatively close to the grids, at distances below approximately $15M$, where $M$ is the grid mesh size and the turbulence may still be spatially inhomogeneous \citep{isaza2014grid,hearst2018robust,kozul2020response}.

Later studies extended the range of Reynolds number and FST intensity. \citet{castro1984effects} investigated momentum-thickness Reynolds number, defined as $Re_\theta = \frac{U_e \theta}{\nu}$, where $U_e$ is the velocity at the boundary-layer edge, $\theta$ is the momentum thickness and $\nu$ is the kinematic viscosity, ranging between $500<Re_\theta<2500$ with turbulence intensities up to approximately $7\%$, showing that both Reynolds number and FST affect skin friction. \citet{blair1983influence1,blair1983influence2} considered $1000<Re_\theta<7000$ and showed that increasing FST increases skin friction and progressively suppresses the wake region and outer-region intermittency, which becomes almost imperceptible for $T_u\simeq5.3\%$. Higher turbulence levels, up to $T_u\simeq20\%$, were later achieved by \citet{thole1996high} using crossflow jets, confirming that wake suppression occurs while the inner-scaled mean velocity remains close to the canonical logarithmic behaviour.

Active-grid experiments provided further control over the intensity and scales of the external turbulence. \citet{sharp2009effects} generated FST levels up to $T_u\simeq10.5\%$ and observed a reduction of wake strength with increasing FST, consistent with earlier studies. \citet{dogan2016interactions} investigated ZPG TBLs subjected to large-scale FST with $T_u\simeq7$-$13\%$, showing that energetic free-stream motions penetrate the boundary layer, increase streamwise velocity fluctuations throughout the boundary layer and produce an energetic outer spectral peak whose magnitude scales with the FST level. The same study showed that the energetic free-stream scales were of order $10\delta$, comparable to the large scales associated with high-Reynolds-number wall turbulence. Subsequent measurements by \citet{dogan2017modelling} further showed that, despite the penetration of FST into the boundary layer, the near-wall region retains several statistical features of canonical high-Reynolds-number TBLs without imposed FST, including enhanced inner-outer interactions and amplitude modulation of near-wall small scales by large outer motions. Direct skin-friction measurements by \citet{esteban2017skin}, performed for FST levels of approximately $2$-$13\%$, confirmed the increase of skin friction under FST and showed good agreement between oil-film interferometry and profile-fitting estimates of the friction velocity.

Most of these studies considered the boundary layer at a fixed streamwise station or under approximately fixed FST conditions. The spatial development of canonical TBLs without imposed FST was documented by \citet{vincenti2013streamwise} and \citet{marusic2015evolution}, who showed that the magnitude of the near-wall variance peak increases as the boundary layer evolves downstream, and that an outer spectral peak emerges with increasing friction Reynolds number. In grid-generated turbulence, however, the TBL grows while the external turbulence decays, so the local state of the flow results from the simultaneous development of both systems. \citet{raushan2018grid} addressed the reciprocal influence between a developing boundary layer and grid-generated free-stream turbulence, using three passive grids in an open water channel and focusing on the evolution of inhomogeneous turbulence in the grid near field. The spatial co-evolution of a turbulent boundary layer and decaying FST was directly investigated by \citet{jooss2021spatial}, who studied TBLs subjected to decaying FST at friction Reynolds numbers up to $Re_\tau = \frac{u_\tau \delta}{\nu}=5060$ and initial turbulence intensities up to $T_u=12.5\%$, where $u_\tau$ is the friction velocity. Their measurements showed that wake suppression is not necessarily permanent: sufficiently far downstream, as the FST decays and the boundary layer continues to grow, the wake region can recover. They also showed that the near-wall variance peak may either grow or decay downstream depending on the initial FST level, highlighting that conventional parameters such as $Re_\tau$ alone are not sufficient to characterize a TBL subjected to decaying FST. Related numerical studies have also confirmed the increase in skin friction and wake suppression under FST. In particular, \citet{you2019conditional} compared a TBL subjected to $T_u\simeq10\%$ with a canonical TBL using DNS at $1900<Re_\theta<3000$, observing an increase in skin friction of up to $15\%$, suppression of the wake region and an increase of the near-wall streamwise variance peak, while the logarithmic region remained unaffected by FST. These observations are particularly relevant when considering more realistic airfoil boundary layers, where the streamwise evolution of the FST occurs simultaneously with the evolution of the pressure gradient imposed by the geometry.

Compared with ZPG TBLs, APG TBLs exhibit substantial changes in both the mean flow and turbulence structure. The pressure-gradient strength is commonly characterized by the Clauser pressure-gradient parameter \citep{monty2011parametric},
\begin{equation}
\beta = \frac{\delta^\ast}{\tau_w}\frac{\mathrm{d}P_e}{\mathrm{d}x},
\end{equation}
where $\delta^\ast$ is the displacement thickness, $\tau_w$ is the wall-shear stress and $P_e$ is the pressure at the boundary-layer edge. APGs strengthen the wake region of the mean velocity profile, increase the shape factor, reduce skin friction and enhance turbulence activity in the outer region of the boundary layer. Previous studies have also identified a secondary peak in turbulence production in the outer region, which can contribute to the redistribution of turbulent energy towards the wall and to the amplification of near-wall normal stresses \citep{vinuesa2017pressure,vinuesa2017revisiting}. Experimental studies by \citet{monty2011parametric,harun2013} and \citet{vila2020experimental} showed that APGs energize large-scale motions in the outer layer, leading to increased streamwise turbulence intensity and stronger inner-outer interaction. Numerical studies of flat-plate APG TBLs have further shown that even mild to moderate APGs at relatively low Reynolds numbers can produce a secondary large-scale peak in the premultiplied spectra of streamwise velocity fluctuations \citep{bobke2017history,lee2017large}. Recent simulations of near-equilibrium APG TBLs at higher Reynolds numbers, including cases with approximately constant $\beta\simeq1.4$ up to $Re_\theta\simeq8700$, showed that APG effects remain associated with outer-layer energization and increased scale separation \citep{pozuelo2022adverse}. However, APG TBLs are also strongly affected by pressure-gradient history and non-equilibrium development, so that local values of $\beta$ and Reynolds number may not be sufficient to characterize the flow state \citep{bobke2017history,volino2020non}.

For airfoils and wings, the pressure gradient is not externally imposed over a flat plate but arises naturally from the geometry and varies along the chord. This introduces curvature-induced acceleration and deceleration, with simultaneous streamwise changes in $\beta$, Reynolds number and boundary-layer thickness. High-fidelity simulations of TBLs developing over airfoil sections have shown that APG effects are strongest in the outer region and that flow history can significantly affect turbulence statistics and spectral organization \citep{tanarro2020effect}. These studies are directly relevant to realistic aerodynamic configurations, but they generally consider clean inflow conditions. Conversely, the FST literature has mostly focused on ZPG TBLs. The combined problem of a spatially developing TBL over an airfoil, simultaneously subjected to APG and elevated FST, therefore remains insufficiently documented.

The interaction between FST and APG effects is not expected to be a simple superposition of the two canonical behaviours. Both mechanisms enhance energetic large-scale motions and influence inner-outer coupling, but they produce opposite trends in several mean-flow quantities. APGs increase the wake deficit, increase the shape factor and reduce skin friction, whereas FST tends to suppress the wake, decrease the shape factor and increase skin friction. Moreover, in a developing airfoil boundary layer, the pressure-gradient history, the local Reynolds number and the local FST level all vary simultaneously. This raises the question of whether FST weakens, enhances or fundamentally modifies the outer-layer response usually associated with APG TBLs.

The present work addresses this gap through wind-tunnel measurements of turbulent boundary layers developing over an airfoil at fixed chord Reynolds number and subjected to controlled levels of grid-generated FST. This configuration is relevant to aerodynamic bodies operating in turbulent environments, such as small airfoils and drones exposed to urban or atmospheric turbulence, where the boundary layer develops under simultaneous pressure-gradient variation and elevated external turbulence. Two angles of attack are considered, producing different APGs, and measurements are performed at four chordwise locations, $x/c=0.400$, $0.475$, $0.550$ and $0.625$, to document the streamwise evolution of the flow. The investigated cases cover local turbulence intensities from approximately $T_u=0.2\%$ to $6\%$, spanning conditions from nearly clean inflow to elevated grid-generated FST, while maintaining the same chord Reynolds number for all cases. The study focuses on the combined influence of FST and APG on the mean-flow development, integral parameters and turbulence structure. Particular attention is given to the extent to which FST modifies the APG-induced wake growth and outer-layer energization, and to whether the response of the boundary layer can be interpreted in terms of the competing effects of free-stream forcing and pressure-gradient-driven development.

The paper is structured as follows. Section~\ref{sec:experiments} describes the wind-tunnel facility, airfoil model, turbulence-generating grids, measurement techniques and characterization of the imposed FST and APG conditions. Section~\ref{sec:results} presents the evolution of the mean-flow quantities, integral parameters and turbulence statistics. The spectral organization of the boundary layer and the scale-dependent contribution of the FST-induced motions are then analysed in Section~\ref{sec:spectra}. Finally, Section~\ref{sec:conclusions} summarizes the main findings and discusses their implications for APG airfoil boundary layers subjected to elevated FST.

\section{Experiments}
\label{sec:experiments}
\subsection{Experimental setup, measurement techniques and procedures}

\subsubsection{Wind tunnel and flow conditions}

\begin{figure}[htbp]
\begin{center}
\includegraphics[width=.9\textwidth]{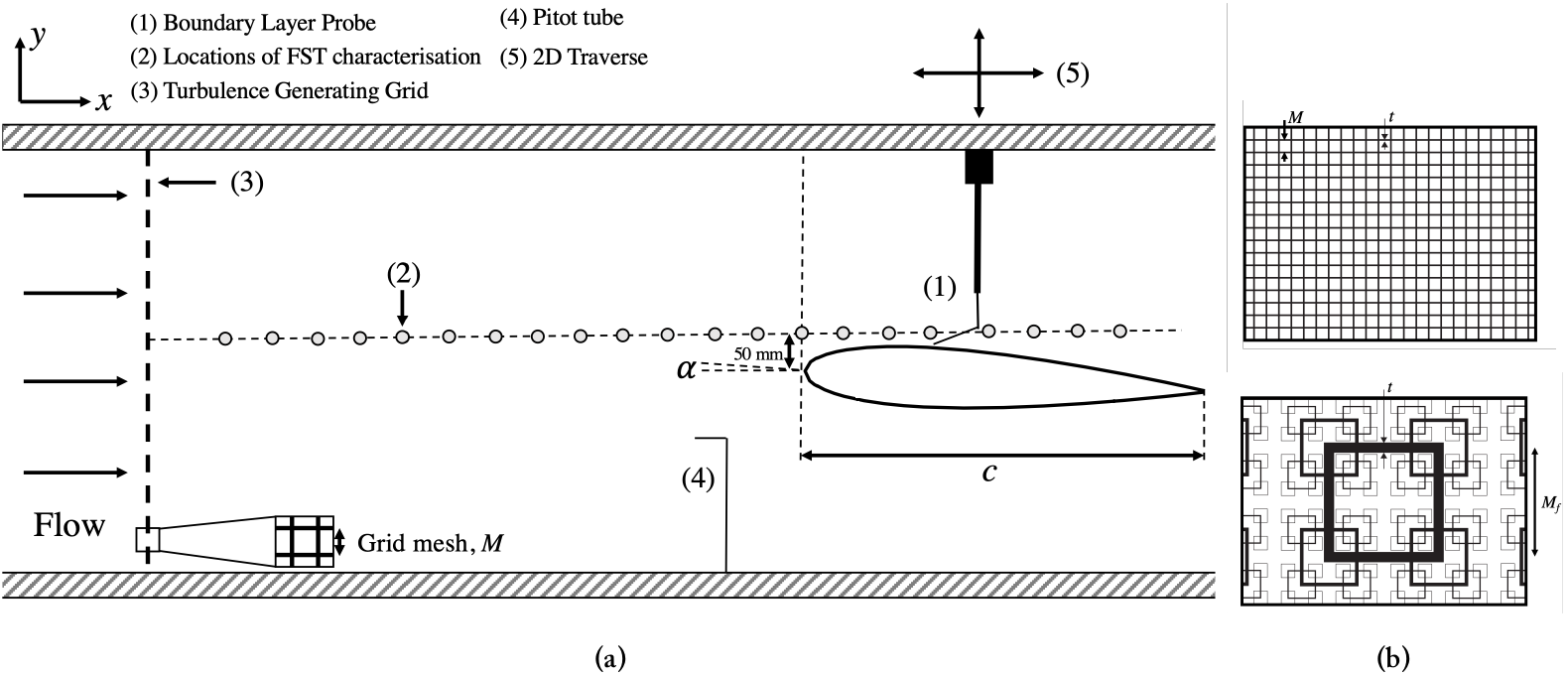}
\end{center}
\caption{Wind tunnel experimental setup (a) and grid samples used to generate free-stream turbulence (b).}
\label{fig:exp_setup_grids}
\end{figure}

The experiments were conducted in the ONERA Toulouse TRIN~2 subsonic wind tunnel, using the same facility, airfoil model, turbulence-generation system and hot-wire arrangement as in \citet{Jaroslawski2023LSB}. The tunnel is operated at atmospheric conditions and has a contraction ratio of 16. The test-section entrance has dimensions of $0.3~\mathrm{m} \times 0.4~\mathrm{m}$ and a total length of $2~\mathrm{m}$. The flow exits the test section through a diffuser and is discharged into a noise-reduction chamber, limiting the propagation of pressure disturbances from the downstream fan into the test section. In the absence of turbulence-generating grids, the free-stream turbulence level measured near the leading edge of the airfoil is around approximately $0.2\%$.

A sketch of the experimental setup is reported in figure~\ref{fig:exp_setup_grids}(a). The model is an aluminium NACA~0015 airfoil with chord length $c=0.3~\mathrm{m}$ and span $0.4~\mathrm{m}$, mounted horizontally in the test section. The leading edge is located $1.44~\mathrm{m}$ downstream of the test-section inlet. The coordinate system is defined with $x$ along the chord, measured from the leading edge, and $y$ normal to the airfoil surface. The chord-based Reynolds number was kept fixed at $Re_c = U_\infty c/\nu \simeq 2.5 \times 10^5$ for all configurations.

Two angles of attack were investigated, $\alpha=\ang{2}$ and $\alpha=\ang{4}$, in order to vary the adverse pressure gradient experienced by the boundary layer while keeping the Reynolds number and incoming turbulence conditions unchanged. The boundary layer on the suction side of the airfoil was tripped upstream of the measurement region to ensure a turbulent state over the investigated chordwise interval. The boundary layer on the suction side of the airfoil was tripped upstream of the measurement region, at x/c = 0.1, to ensure a turbulent state over the investigated chordwise interval. The trip consisted of a 1 mm wire applied over the full span of the model.

\subsubsection{Grid-generated free-stream turbulence}

Free-stream turbulence was generated using passive turbulence-generating grids mounted upstream of the airfoil, as shown in figure~\ref{fig:exp_setup_grids}(b). The grid configurations are the same as those employed by \citet{Jaroslawski2023LSB}, and include regular grids and a fractal grid designed to provide a various of turbulence intensities. The grid geometry is characterized by the mesh size $M$, bar thickness $t$ and solidity $\sigma$. The grids were positioned at least $20M$ upstream of the leading edge, allowing the generated turbulence to develop before interacting with the airfoil boundary layer.

The free-stream turbulence was characterized using X-wire (Dantec 55P51 X-wire on a 55H24 support) measurements at locations outside the boundary layer. Measurements were performed along the wind-tunnel centreline upstream of the airfoil and above the suction side of the airfoil, following the procedure used by \citet{Jaroslawski2023LSB}. For each grid configuration, records of approximately 120 s, sampled at 25 kHz, were used to ensure statistical convergence of the turbulence quantities, corresponding to at least 40 000–60 000 integral length scales of the flow. The turbulence intensity, integral length scale and spectral content of the external flow were evaluated at the same chordwise stations as the boundary-layer profiles, with the probe positioned outside the boundary layer.

Four incoming turbulence conditions are considered in the present study depending on the turbulence intensity. The streamwise turbulence intensity is defined as
\begin{equation}
    T_u = \frac{u_{\mathrm{rms}}}{U_\infty},
\end{equation}
where $u_{\mathrm{rms}}$ is the root-mean-square value of the streamwise velocity fluctuations measured outside the boundary layer.
The lowest-turbulence case corresponds to the clean wind natural tunnel condition, with $T_u \simeq 0.2\%$. The remaining cases are generated using static grids and correspond to turbulence levels of approximately $T_u=1.4\%$, $2.5\%$ and $6.0\%$ in the region of interest. The same grid configurations were used for both angles of attack, so that the effect of the pressure gradient could be investigated at nominally fixed incoming turbulence conditions.

\subsubsection{Turbulent boundary layer measurements}

Boundary-layer measurements were performed using a single-component hot-wire boundary-layer probe (Dantec 55P15), operated with a Dantec Dynamics Streamline Pro constant-temperature anemometer (90C10 module) and mounted on a two-dimensional traverse whose position was measured using Heidenhain LS388 linear encoders with a stepping accuracy of 5 µm. Signals were acquired with a National Instruments CompactDAQ-9178 chassis (two NI-9239 24-bit voltage modules and a NI-9211 temperature module) at a sampling frequency of 25 kHz, with an anti-aliasing filter applied by the acquisition card. The system frequency response, estimated from a standard pulse-response test, was approximately 45 kHz. Mean and fluctuating profiles were acquired for 10 s at each wall-normal location, ensuring convergence of second-order statistics to within ±1 \% (95 \% confidence interval), with a 10 s stabilization time imposed after each traverse displacement. The closest measurements to the wall were taken at 200 µm to avoid near-wall thermal corrections. The positioning uncertainty was below 0.05 mm and the hot-wire measurement uncertainty was estimated to be less than 3\% for $U/U_{\infty} > 0.2$.

Wall-normal profiles of the mean and fluctuating streamwise velocity were acquired at four chordwise locations on the suction side of the airfoil,
\begin{equation}
    x/c = 0.400,\;0.475,\;0.550,\;0.625 .
\end{equation}
These locations cover the spatial development of the turbulent boundary layer over the region of pressure recovery on the suction side.

The friction velocity $u_\tau$ was obtained by fitting the measured mean velocity profile using the method of \citet{rodriguez-lopez_robust_2015}. This approach avoids relying directly on near-wall velocity gradients, which are difficult to evaluate accurately with finite-size hot-wire probes and moderate wall-normal resolution. The method has been employed for both adverse-pressure-gradient turbulent boundary layers and zero-pressure-gradient turbulent boundary layers subjected to free-stream turbulence \citep{jooss2021spatial}.
The resulting values of $u_\tau$ range from approximately $0.61$ to $0.86~\mathrm{m\,s^{-1}}$ across the investigated cases. Based on these values, the hot-wire sensing length corresponds to $\ell^+ \simeq 48$--$68$. The finite sensing length is therefore expected to attenuate the measured small-scale structures, in particular near the wall, but not to affect the large-scale motions and outer-layer trends discussed in the following sections.

The parameters usefull to identify the flow regimes in terms of the boundary layer quantities, including $Re_\theta = U_e\theta/\nu$ and $Re_\tau = u_\tau\delta/\nu$, are summarized in table~\ref{tab:parameters}.

\subsection{Characterisation of flow conditions: free-stream turbulence and adverse pressure gradient}

\subsubsection{Adverse pressure gradient over the airfoil}

\begin{figure}[htbp]
\begin{center}
\includegraphics[width=1\textwidth]{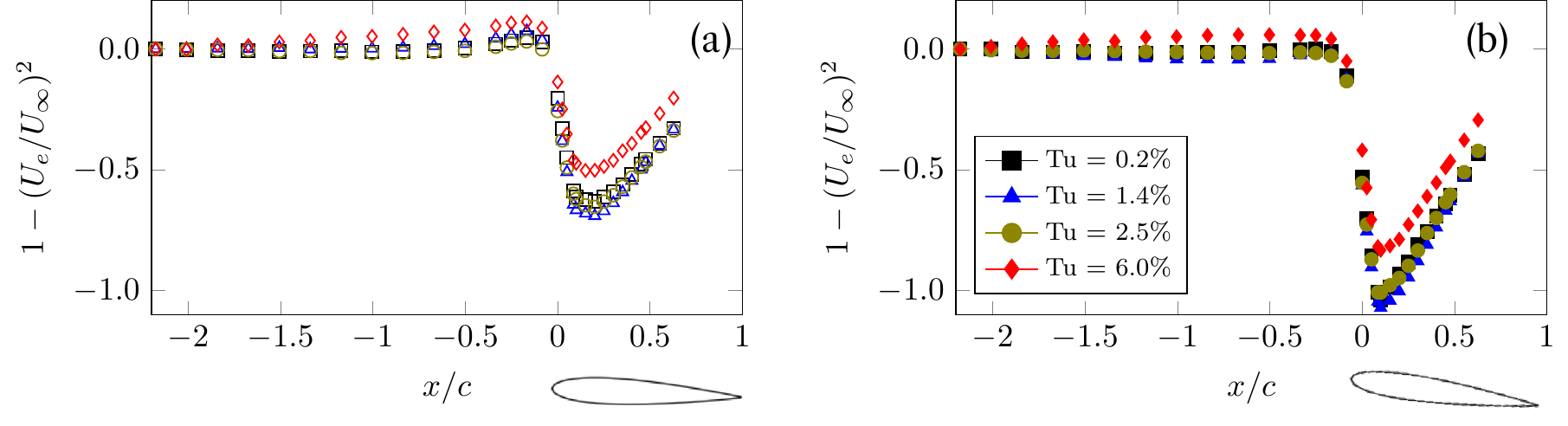}
\end{center}
\caption{Pressure coefficient evolution in the wind tunnel and over the airfoil (a) AoA of \ang{2}, (b) AoA of \ang{4}. }
\label{fig:Cp}
\end{figure}

The pressure coefficient over the airfoil is shown in figure~\ref{fig:Cp} for the two angles of attack and the four free-stream turbulence levels. The pressure coefficient is defined as
\begin{equation}
    C_p = 1 - \left(\frac{U_e}{U_\infty}\right)^2 ,
\end{equation}
where $U_e$ is the local edge velocity. For both angles of attack, the boundary-layer measurement region is located downstream of the suction peak, where the flow experiences a streamwise deceleration and therefore an adverse pressure gradient. Increasing the angle of attack from $\ang{2}$ to $\ang{4}$ strengthens the suction peak and increases the subsequent pressure recovery, resulting in a stronger adverse pressure gradient over the measurement region.

For the three lowest turbulence levels, the pressure distribution remains close to the clean configuration, indicating that the imposed FST do not substantially modify the outer inviscid flow. A noticeable modification is observed only for the highest turbulence level, $T_u \simeq 6\%$, suggesting that the strong external forcing modifies the boundary-layer development sufficiently to affect the pressure recovery. This behaviour is observed for both angles of attack.

The pressure-gradient strength is quantified using the Clauser pressure-gradient parameter,
\begin{equation}
    \beta = \frac{\delta^\ast}{\tau_w}\frac{\mathrm{d}P_e}{\mathrm{d}x},
\end{equation}
where $P_e$ is the boundary-layer edge pressure, $\delta^\ast$ is the displacement thickness and $\tau_w=\rho u_\tau^2$ is the wall-shear stress. In the present database, $\beta$ ranges from approximately $0.3$ to $1.5$, corresponding to a moderate adverse-pressure-gradient regime.

\begin{figure}[htbp]
\begin{center}
\includegraphics[width=.7\textwidth]{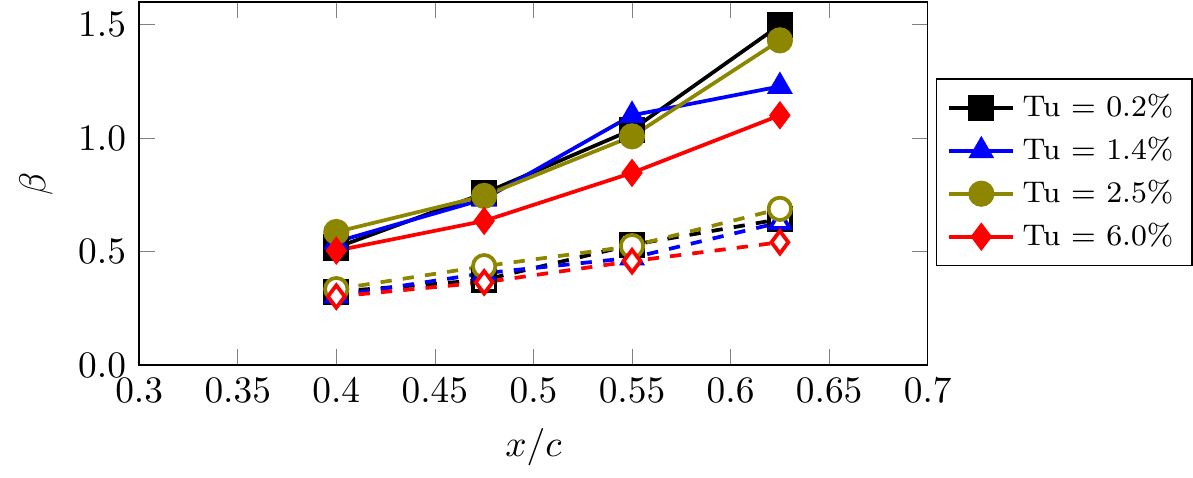}
\end{center}
\caption{Pressure gradient parameter $\beta$ for the four measurement location over the airfoil, for the four $T_u$ tested, open marks represent AoA of \ang{2}, solid marks represent AoA of \ang{4}. }
\label{fig:beta_xc}
\end{figure}

The streamwise evolution of $\beta$ is reported in figure~\ref{fig:beta_xc}. For all free-stream turbulence levels, $\beta$ increases with downstream distance, reflecting the combined effect of boundary-layer growth and pressure recovery along the suction side of the airfoil. At a fixed chordwise location, the $\ang{4}$ cases exhibit larger values of $\beta$ than the $\ang{2}$ cases, confirming that the angle of attack provides an effective way of varying the adverse pressure gradient while keeping the same airfoil geometry and chord-based Reynolds number.

A dependence of $\beta$ on the free-stream turbulence level is also observed. This dependence arises because $\beta$ contains both outer-flow information, through $\mathrm{d}P_e/\mathrm{d}x$, and boundary-layer quantities, through $\delta^\ast$ and $\tau_w$. As shown later, increasing $T_u$ modifies the boundary-layer integral properties and wall shear, and therefore also changes the resulting value of $\beta$.

\subsubsection{Free-stream turbulence characterisation}

\begin{figure}[htbp]
\begin{center}
\includegraphics[width=1\textwidth]{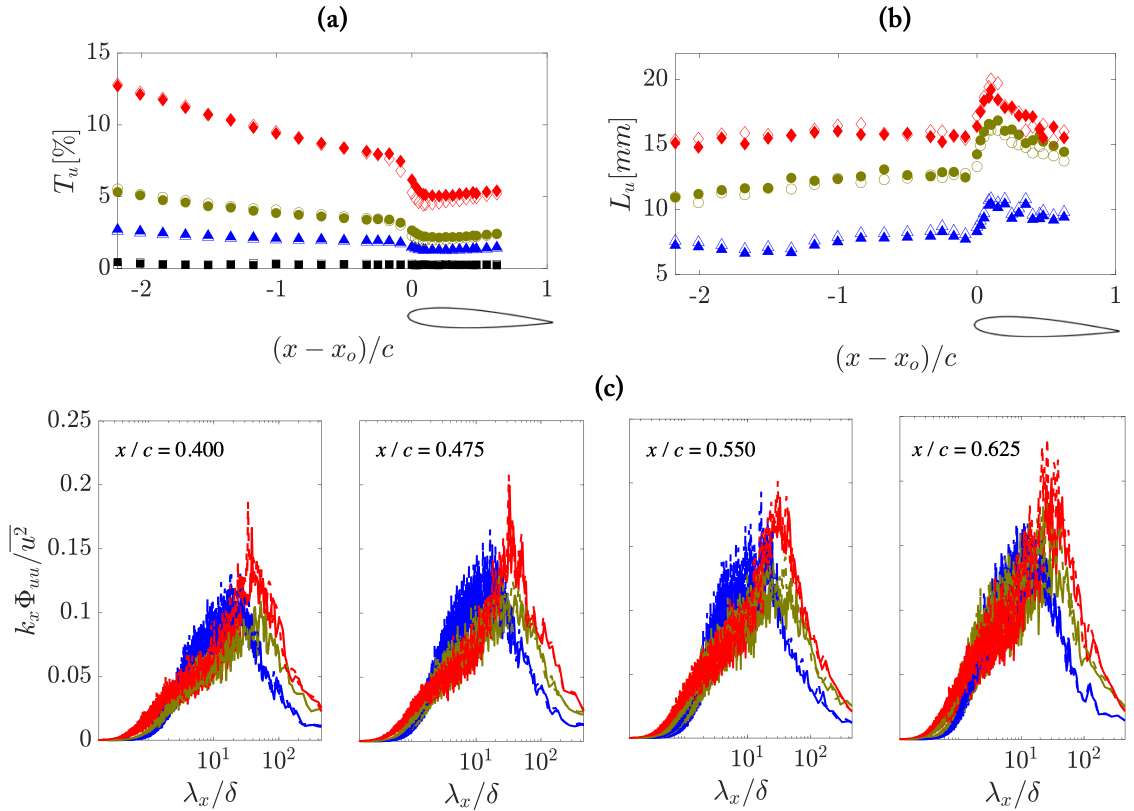}
\end{center}
\caption{Turbulence intensity evolution (a) and integral length-scale (b) in the wind tunnel and over the airfoil, colours and symbols as in \fref{fig:Cp}, open marks represent AoA of \ang{2}, solid marks represent AoA of \ang{4}; (c) premultiplied spectra at the four measurement locations external to the boundary layer.}
\label{fig:Tu_characterisation}
\end{figure}

The free-stream turbulence conditions are characterized in figure~\ref{fig:Tu_characterisation}. Figure~\ref{fig:Tu_characterisation}(a) shows the streamwise evolution of $T_u$ upstream of and over the airfoil. The clean tunnel condition remains close to $T_u\simeq0.2\%$, while the grid-generated cases provide turbulence levels of approximately $1.4\%$, $2.5\%$ and $6.0\%$ over the measurement region. The decay of $T_u$ with streamwise distance is more pronounced for the highest turbulence level. Over the airfoil, a slight variation of Tu is observed. Unlike zero-pressure-gradient boundary layers, in which Tu continues to decay in the streamwise direction, here Tu remains rather constant over most of the airfoil and decreases only near the leading edge for the highest-Tu configurations. This behaviour is attributed to the favourable pressure gradient near the leading edge of the airfoil, which locally accelerates the flow and offsets the streamwise decay of the grid-generated turbulence. No major differences are observed when changing the angle of attack, except for moderate deviations at the highest turbulence level. The two angles of attack can therefore be considered as approximately iso-$T_u$ configurations.

The streamwise integral length scale $L_u$ is obtained from the autocorrelation of the streamwise velocity fluctuation signal, using Taylor's frozen-turbulence hypothesis,
\begin{equation}
    L_u = U_\infty \int_0^{\tau_0} R_{uu}(\tau)\,\mathrm{d}\tau ,
\end{equation}
where $R_{uu}$ is the temporal autocorrelation coefficient and $\tau_0$ is taken as the first zero crossing of the autocorrelation function. This procedure follows that used in the previous characterization of the same turbulence-generating grids by \citet{jaroslawski2023boundary}.
The streamwise integral length scale is reported in figure~\ref{fig:Tu_characterisation}(b). The length scale remains approximately constant upstream of the airfoil, then varies over the suction side. This evolution is attributed to the increased velocity near the leading edge of the airfoil, which stretches the energy-containing free-stream eddies and produces a modest increase in the streamwise integral length scale over the suction side. A similar behaviour was reported in the same facility for the same turbulence-generating grids by \cite{Jaroslawski2023LSB}

Premultiplied spectra measured outside the boundary layer are shown in figure~\ref{fig:Tu_characterisation}(c) at the four chordwise locations. For conciseness, only the $\alpha=\ang{4}$ case is reported. The wavelength is obtained using Taylor's hypothesis and normalized by a reference boundary-layer thickness, $\delta$,taken as the value for the clean configuration at the first measurement location. The energetic content of the free-stream turbulence is centred at wavelengths of approximately $10$-$20\delta$, indicating that the imposed free-stream turbulence contains scales substantially larger than the local boundary-layer thickness. The results are similar with what reported by \citet{dogan2016interactions,dogan2017modelling}. Increasing $T_u$ leads to a systematic increase in external spectral energy over a broad wavelength range, as expected.
A moderate spectral shift is observed for the highest turbulence level, this can suggest that the most energetic wavelength tends to increase with the FST. 

Overall, the flow characterization confirms that the present database combines two controlled effects: increasing free-stream turbulence intensity through passive grids, and increasing adverse pressure gradient through the angle of attack and downstream development over the airfoil.

\begin{table}[htbp]
\centering
\caption{Summary of local boundary-layer parameters for all investigated cases. }
\label{tab:parameters}
\small
\begin{tabular}{ccccc@{\qquad}ccccc}
\toprule
\multicolumn{5}{c}{$\alpha=\ang{2}$} 
& \multicolumn{5}{c}{$\alpha=\ang{4}$} \\
\cmidrule(lr){1-5}\cmidrule(lr){6-10}
Symbol & $x/c$ & $T_u~[\%]$ & $Re_\theta$ & $Re_\tau$ & 
Symbol & $x/c$ & $T_u~[\%]$ & $Re_\theta$ & $Re_\tau$ \\
\midrule

\textcolor{black}{$\square$} & 0.400 & 0.3 & 613 & 267
& \textcolor{black}{$\blacksquare$} & 0.400 & 0.3 & 806 & 308 \\

\textcolor{black}{$\square$} & 0.475 & 0.3 & 706 & 285
& \textcolor{black}{$\blacksquare$} & 0.475 & 0.3 & 965 & 319 \\

\textcolor{black}{$\square$} & 0.550 & 0.4 & 836 & 288
& \textcolor{black}{$\blacksquare$} & 0.550 & 0.4 & 1147 & 331 \\

\textcolor{black}{$\square$} & 0.625 & 0.5 & 964 & 312
& \textcolor{black}{$\blacksquare$} & 0.625 & 0.5 & 1455 & 344 \\
\cmidrule(lr){1-5}\cmidrule(lr){6-10}

\textcolor{blue}{$\triangle$} & 0.400 & 1.4 & 598 & 269
& \textcolor{blue}{$\blacktriangle$} & 0.400 & 1.4 & 756 & 296 \\

\textcolor{blue}{$\triangle$} & 0.475 & 1.4 & 726 & 304
& \textcolor{blue}{$\blacktriangle$} & 0.475 & 1.4 & 887 & 316 \\

\textcolor{blue}{$\triangle$} & 0.550 & 1.6 & 815 & 300
& \textcolor{blue}{$\blacktriangle$} & 0.550 & 1.6 & 1163 & 326 \\

\textcolor{blue}{$\triangle$} & 0.625 & 1.7 & 913 & 305
& \textcolor{blue}{$\blacktriangle$} & 0.625 & 1.7 & 1101 & 297 \\
\cmidrule(lr){1-5}\cmidrule(lr){6-10}

\textcolor{olive}{$\circ$} & 0.400 & 2.5 & 605 & 275
& \textcolor{olive}{$\bullet$} & 0.400 & 2.5 & 777 & 286 \\

\textcolor{olive}{$\circ$} & 0.475 & 2.6 & 709 & 296
& \textcolor{olive}{$\bullet$} & 0.475 & 2.6 & 931 & 336 \\

\textcolor{olive}{$\circ$} & 0.550 & 2.8 & 804 & 302
& \textcolor{olive}{$\bullet$} & 0.550 & 2.8 & 1105 & 359 \\

\textcolor{olive}{$\circ$} & 0.625 & 3.1 & 928 & 316
& \textcolor{olive}{$\bullet$} & 0.625 & 3.1 & 1304 & 392 \\
\cmidrule(lr){1-5}\cmidrule(lr){6-10}

\textcolor{red}{$\diamond$} & 0.400 & 5.3 & 713 & 343
& \textcolor{red}{$\blacklozenge$} & 0.400 & 5.6 & 897 & 416 \\

\textcolor{red}{$\diamond$} & 0.475 & 5.6 & 829 & 381
& \textcolor{red}{$\blacklozenge$} & 0.475 & 5.8 & 1016 & 395 \\

\textcolor{red}{$\diamond$} & 0.550 & 6.0 & 949 & 395
& \textcolor{red}{$\blacklozenge$} & 0.550 & 6.1 & 1117 & 416 \\

\textcolor{red}{$\diamond$} & 0.625 & 6.4 & 1044 & 396
& \textcolor{red}{$\blacklozenge$} & 0.625 & 6.5 & 1299 & 442 \\

\bottomrule
\end{tabular}
\end{table}

\section{Results}
\label{sec:results}
\subsection{Boundary layer parameters}

The boundary-layer thickness $\delta$ was evaluated following the diagnostic-profile method proposed by \citet{rodriguez-lopez_robust_2015}, rather than the method used by \citet{vinuesa_determining_2016}. This choice was made because the present profiles are affected by both adverse pressure gradient and free-stream turbulence, making the threshold definition based on the diagnostic plot sensitive to the external turbulence level and to the precise identification of the boundary-layer edge (see \fref{fig:diagnostic} in Appendix \ref{sec:appendixA}).

\begin{figure}[htbp]
\begin{center}
\includegraphics[width=.9\textwidth]{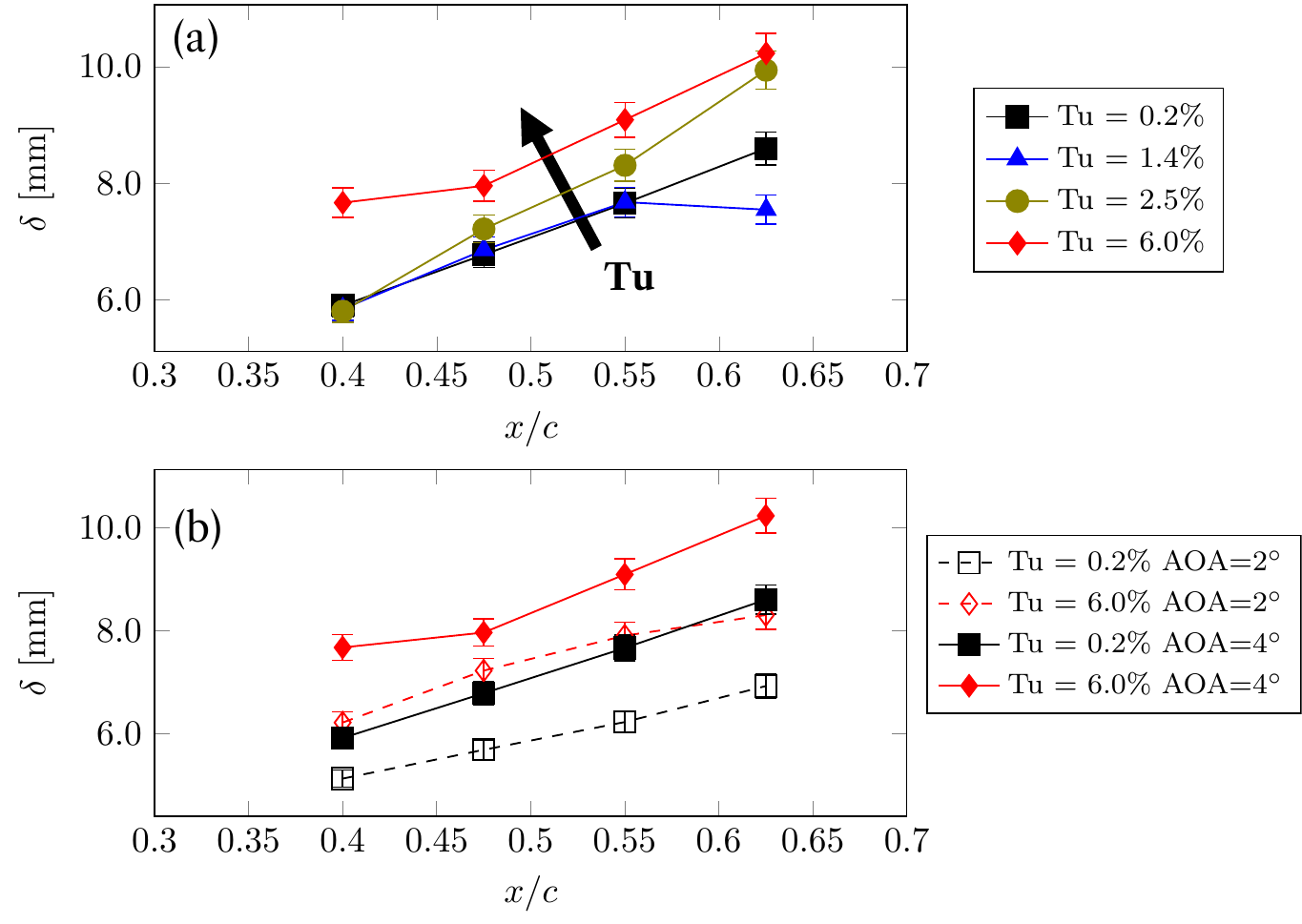}
\end{center}
\caption{Boundary layer thickness as function of the streamwise coordinate $x/c$ along the airfoil, (a) effect of changing the Tu at an AoA = \ang{4}, (b) effect of changing the AoA for the lowest and highest Tu.}
\label{fig:delta_streamwise}
\end{figure}

The streamwise evolution of the boundary layer thickness $\delta$ is shown in \fref{fig:delta_streamwise}. In \fref{fig:delta_streamwise}(a), results are reported for a fixed angle of attack of $\ang{4}$, corresponding to the most severe adverse pressure gradient, for four different freestream turbulence intensities $T_u$. Since the flow is spatially developing, the boundary layer exhibits a monotonic growth when moving downstream, from $x/c = 0.4$ to $x/c = 0.625$. In addition to the natural streamwise development of the turbulent boundary layer, the pressure-gradient parameter $\beta$ increases with streamwise position, leading to a combined effect of boundary layer growth and progressively stronger adverse pressure gradients.

For the lowest freestream turbulence level, the boundary layer thickness increases from approximately $6\,\mathrm{mm}$ to $8.2\,\mathrm{mm}$ over the measurement region, corresponding to a relative increase of about $\sim 35\%$. A similar relative growth is observed for the other freestream turbulence conditions. While the two lowest turbulence levels ($T_u \approx 0.2\%$ and $1.4\%$) yield nearly overlapping thickness distributions, indicating a minor influence of freestream turbulence, a clear increase in $\delta$ is observed when the freestream turbulence intensity is further increased. As highlighted by the arrow in \fref{fig:delta_streamwise}(a), elevated freestream turbulence levels systematically lead to a thicker boundary layer, despite the decreasing trend of $\beta$ at a fixed streamwise location.

The effect of the pressure gradient is further illustrated in \fref{fig:delta_streamwise}(b), where results for the lowest and highest freestream turbulence levels are shown for two angles of attack. The dashed lines correspond to an angle of attack of $\ang{2}$, while the solid lines represent $\ang{4}$. For both freestream turbulence conditions, increasing the angle of attack and thus strengthening the adverse pressure gradient, as shown in \fref{fig:beta_xc} results in an overall increase of the boundary layer thickness along the airfoil.

Overall, these results indicate that freestream turbulence promotes boundary layer thickening in spatially developing adverse-pressure-gradient turbulent boundary layers, acting as an external disturbance that enhances boundary layer growth in a manner qualitatively similar to that observed in spatially developing ZPG turbulent boundary layers subjected to freestream turbulence \citep{jooss2021spatial,dogan2016interactions}.

\begin{figure}[htbp]
\begin{center}
\includegraphics[width=.9\textwidth]{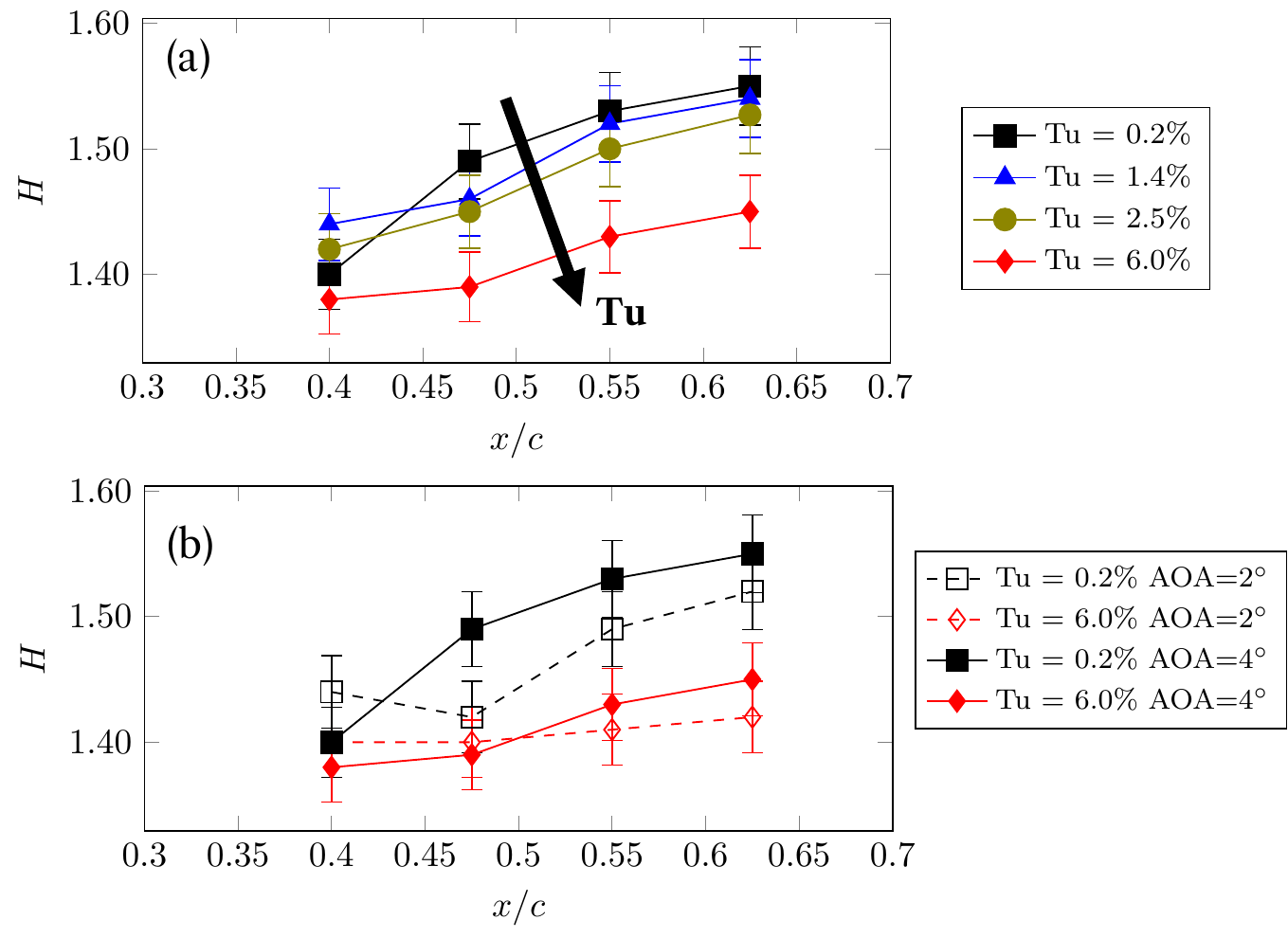}
\end{center}
\caption{Shape factor as function of the streamwise coordinate $x/c$ along the airfoil, (a) effect of changing the Tu at an AoA = \ang{4}, (b) effect of changing the AoA for the lowest and highest Tu.}
\label{fig:H_streamwise}
\end{figure}

The shape factor $H$, defined as the ratio between the displacement thickness and the momentum thickness is reported in \fref{fig:H_streamwise}. Again the subfigure (a) is at AoA of 4 deg showing the effect of the fst at various x/c. Since the TBL is evolving and when moving downstream beta increases the shape factor increases downstream due to a more severe adverse pressure gradient. The effect of the Tu is to decrease the shape factor as indicated by the arrow. If for low Tu the values are betwen 1.45 and 1.55 when introducing the highest FST thet decrease to 1.4 to 1.45 approzimately
Low fst the results clearly show the trend but the varuations is within the uncertainty of the measurements.
The effect is similar to what happens for ZPG turbulent boundary layers \citep{blair1983influence2,jooss2021spatial}.

The subfigure (b) evidences the effect of changing the AoA for the lowest and highest $T_u$. AoA, thus increasing beta,  increases the shape factor as expected \citep{monty2011parametric,harun2013,vila2020experimental}, and this effect is more pronunced for low value of $T_u$ while for the highest Tu the value of shape factor are almost identical with the shape factor that almost flattens. This evidences a predominant effect of the FST over the beta and the shape factor assumes a value of 1.4 typical of a ZPG boundary layer, despide the presence of APG.

The results indicate that the the external fst leads to an increase of momentum within the boundary layer  suggesting a penetration of the energy of the freestream within the TBL that compensate the effect of the TBL that is to increase the shape factor.

\begin{figure}[htbp]
\begin{center}
\includegraphics[width=1\textwidth]{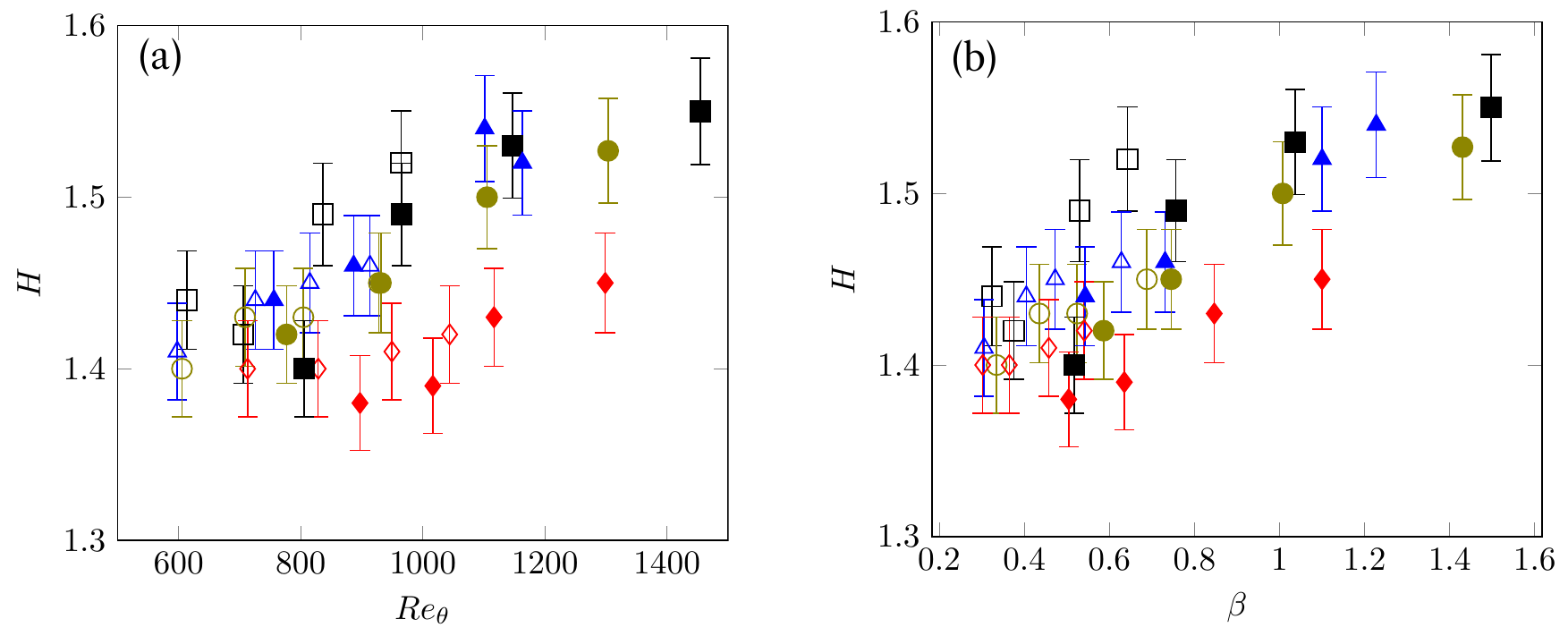}
\end{center}
\caption{Shape factor as function of momentum thickness Reynolds number (a) and pressure gradient parameter $\beta$, colours and symbols as in \fref{fig:Cp}, colours and symbols as in \fref{fig:Cp}, open marks represent AoA of \ang{2}, solid marks represent AoA of \ang{4}.}
\label{fig:H_Rethetabeta}
\end{figure}

In \fref{fig:H_Rethetabeta}, the shape factor $H$ is reported as a function of (a) the momentum-thickness Reynolds number $Re_{\theta}$ and (b) the pressure-gradient parameter $\beta$. These results should be interpreted with caution, as each measurement reflects the combined effects of spatial boundary layer development and varying adverse pressure gradients, preventing a strict decoupling of their individual contributions.

For the lowest freestream turbulence level, a clear increase in the shape factor is observed with increasing $Re_{\theta}$ and $\beta$. At low values of $\beta$, corresponding to negligible adverse pressure gradients, the measured shape-factor values are comparable to those reported for canonical ZPG turbulent boundary layers \citep{schlatter2010,FERNHOLZ1996245}. 
In contrast, for the highest freestream turbulence level, the shape factor remains significantly lower than in the low-$T_u$ case. Although a mild increase in $H$ is observed for $\beta \gtrsim 0.8$ and $Re_{\theta} \gtrsim 1000$, this trend remains within the experimental uncertainty. The deviation between low and high freestream turbulence conditions becomes increasingly pronounced as either $Re_{\theta}$ or $\beta$ increases, whereas at low $\beta$ the shape-factor values collapse for all freestream turbulence levels.

These observations indicate that the influence of freestream turbulence on the shape factor becomes more significant at higher Reynolds numbers and stronger adverse pressure gradients. This behavior suggests that, under increasing APG conditions, freestream turbulence may further enhance momentum transport towards the near-wall region of the turbulent boundary layer, thereby reducing the shape factor.

\subsection{Skin friction coefficient}

\begin{figure}[htbp]
\begin{center}
\includegraphics[width=.5\textwidth]{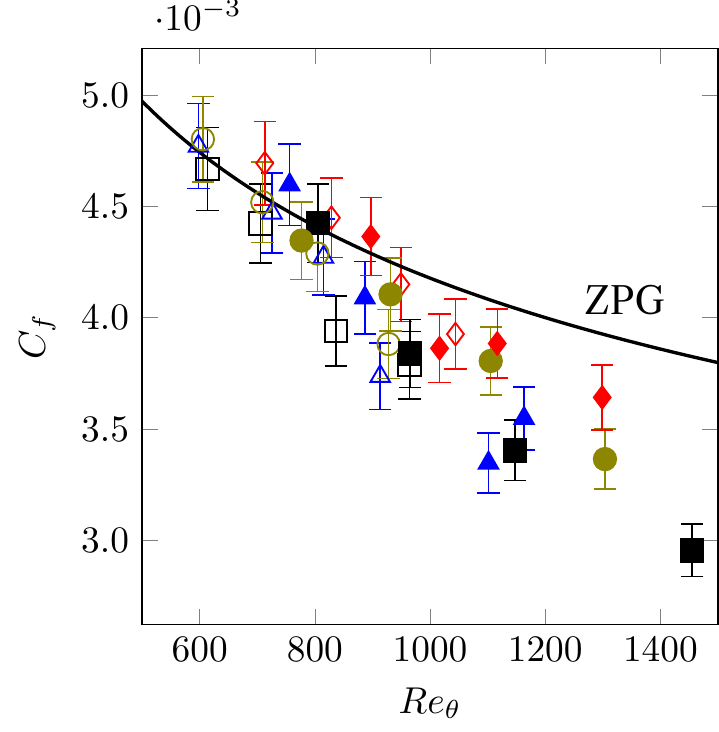}
\end{center}
\caption{Skin friction coefficient as function of momentum thickness Reynolds number, colours and symbols as in \fref{fig:Cp}, open marks represent AoA of \ang{2}, solid marks represent AoA of \ang{4}. The solid line represents Coles-Fernholz relation \citep{FERNHOLZ1996245} for zero pressure gradient turbulent boundary layers.}
\label{fig:Cf_Retheta}
\end{figure}

In \fref{fig:Cf_Retheta}, the skin-friction coefficient $C_f$ is shown as a function of the momentum-thickness Reynolds number $Re_{\theta}$. 
% The parameter $Re_{\theta}$ can be regarded as a proxy for the spatial development of the turbulent boundary layer along the airfoil and is directly linked to the pressure-gradient parameter $\beta$. Consequently, variations in $Re_{\theta}$ also provide qualitative insight into the influence of the adverse pressure gradient, as further illustrated in Appendix~\fref{fig:beta_Retheta}.
For reference, the solid line in \fref{fig:Cf_Retheta} represents the Coles–Fernholz relation for zero-pressure-gradient turbulent boundary layers \citep{FERNHOLZ1996245}. At low values of $Re_{\theta}$ (and thus weak adverse pressure gradients), the measured skin-friction coefficients closely follow the ZPG correlation. As $Re_{\theta}$ increases, a clear reduction in $C_f$ is observed for the lowest freestream turbulence levels, consistent with the expected behavior of turbulent boundary layers subjected to adverse pressure gradients \citep{monty2011parametric,harun2013,tanarro2020effect,pozuelo2022adverse}.

In contrast, for the highest freestream turbulence condition, the skin-friction coefficient increases with $Re_{\theta}$ and progressively approaches the ZPG reference curve. This behavior suggests that, in the presence of strong freestream turbulence, the effect of the adverse pressure gradient on the turbulent boundary layer is partially mitigated. The additional momentum transport towards the near-wall region induced by freestream turbulence may lead to an increase in wall shear stress, resulting in higher skin-friction levels despite the strengthening adverse pressure gradient, consistent with previous ZPG studies under FST \citep{blair1983influence1,esteban2017skin,you2019conditional}.

\subsection{Mean and variance profiles}

\begin{figure}[htbp]
\begin{center}
\includegraphics[width=.9\textwidth]{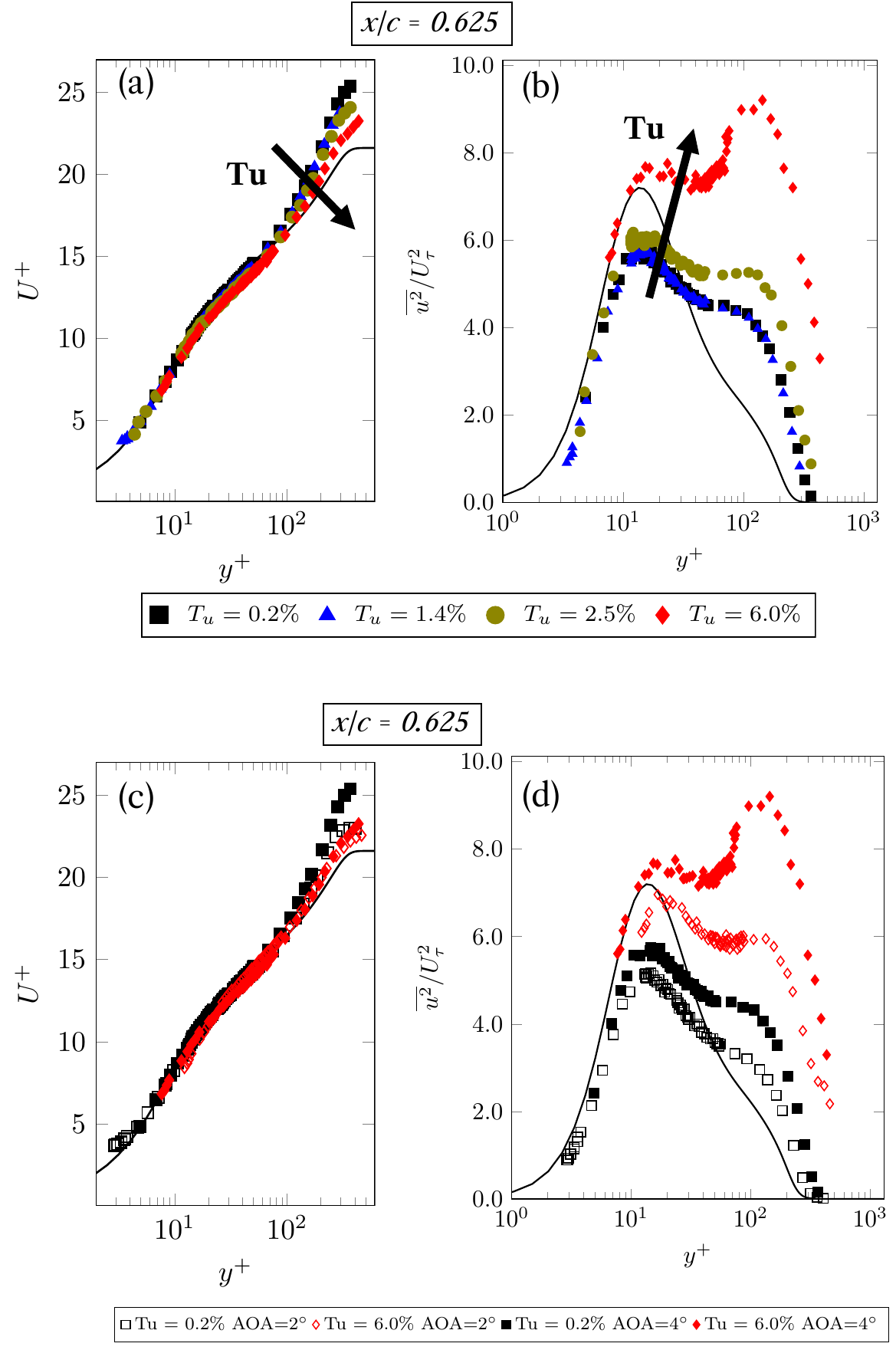}
\end{center}
\caption{Profiles of mean (a,c) and variance (b,d) of the streamwise velocity in wall units at $x/c=0.625$; (a,b) effect of changing the $T_u$ at $AOA=\ang{4}$, (c,d) effect of changing the AoA for the lowest and highest $T_u$, open marks represent AoA of \ang{2}, solid marks represent AoA of \ang{4}.}
\label{fig:profiles_Tu_AoA}
\end{figure}

Figure~\ref{fig:profiles_Tu_AoA} presents the wall-normal profiles of the mean streamwise velocity (a,c) and the corresponding streamwise velocity variance (b,d), compared with numerical data from \cite{schlatter2010} at comparable values of $Re_{\theta}$. 
Results in panels (a,b) show the results when varying the $T_u$ for a fixed streamwise location, $x/c = 0.6235$, and at the highest angle of attack, corresponding to the most severe condition investigated here, where the combined effects of boundary-layer development and adverse pressure gradient are strongest (see \fref{fig:beta_xc}).
Panels (c,d) illustrate the effect of varying the angle of attack at the lowest and highest freestream turbulence levels.

The mean velocity profile at $T_u=0.2\%$ exhibits a pronounced wake region. This behavior is characteristic of APG turbulent boundary layers and reflects the growing influence of the pressure gradient on the outer region of the flow. 
Increasing the angle of attack from \ang{2} to \ang{4} strengthens the adverse pressure gradient, thus increasing the value of $\beta$ leads to a more pronounced wake in the mean velocity profile, as shown in \fref{fig:profiles_Tu_AoA} (c).
This result is consistent with previous observations in APG turbulent boundary layers \citep{monty2011parametric,harun2013,vila2020experimental,pozuelo2022adverse}.

However, increasing freestream turbulence intensity leads to a systematic reduction of the wake strength. This trend mirrors the wake suppression reported for ZPG turbulent boundary layers subjected to freestream turbulence \citep{blair1983influence2,sharp2009effects,dogan2016interactions,jooss2021spatial}, suggesting that elevated freestream turbulence promotes enhanced mixing in the outer region of the boundary layer. The results at high value of freestream turbulence approach the canonical boundary layer under ZPG reported as reference.

The logarithmic region of the mean velocity profiles does not exhibit significant variations in slope across the investigated conditions. Previous studies have reported a decrease of the von K\'arm\'an constant $\kappa$ in APG turbulent boundary layers, leading to steeper mean velocity profiles compared to ZPG cases \citep{vinuesa2017pressure,vinuesa2017revisiting}. In the present experiments, no clear modification of $\kappa$ is observed, likely due to the moderate strength of the imposed adverse pressure gradients.

The corresponding streamwise velocity variance profiles are shown in \fref{fig:profiles_Tu_AoA}(b,d). For the Reynolds numbers investigated here, canonical ZPG turbulent boundary layers would not be expected to exhibit a pronounced outer peak in the profile variance \cite{Vallikivi2015TurbulentNumber}. Consistently, the reference ZPG data display only a near-wall peak. The inner peak measured in the present experiments is partially attenuated due to spatial filtering associated with the hot-wire sensor length, explaining the mismatch with the numerical data close to the wall.

The presence of an adverse pressure gradient promotes the emergence of an outer peak in the variance profile \citep{monty2011parametric,harun2013,bobke2017history,lee2017large}. 
Increasing the angle of attack (\fref{fig:profiles_Tu_AoA} (d)) leads to an amplification of both the inner and outer peaks in the streamwise velocity variance. 
The outer peak is already visible at low freestream turbulence levels and becomes enhanced as the freestream turbulence intensity increases (\fref{fig:profiles_Tu_AoA} (b)). Simultaneously, an increase in the inner peak magnitude is observed. These trends indicate that freestream turbulence penetrates the turbulent boundary layer and amplifies turbulent activity across both inner and outer regions as shown by previous works \citep{dogan2016interactions,dogan2017modelling,jooss2021spatial,you2019conditional}. Notably, at this streamwise location the value of $\beta$ corresponding to $T_u = 6\%$ is lower than that associated with the highest adverse pressure gradient condition, suggesting that the enhanced outer peak can be attributed primarily to the additional turbulent energy introduced by the freestream rather than to pressure-gradient effects alone.

These results further confirm the combined role of adverse pressure gradients and freestream turbulence in shaping the turbulence structure of spatially developing boundary layers.
The simultaneous presence of an outer peak in the variance and the profile which approaches the one of the canonical under ZPG might indicate that the combined presence of APG and moderate value of $T_u$ reproduces some features commonly associated with higher-Reynolds-number turbulent boundary layers \citep{dogan2017modelling}.

\begin{figure}[htbp]
\begin{center}
\includegraphics[width=1\textwidth]{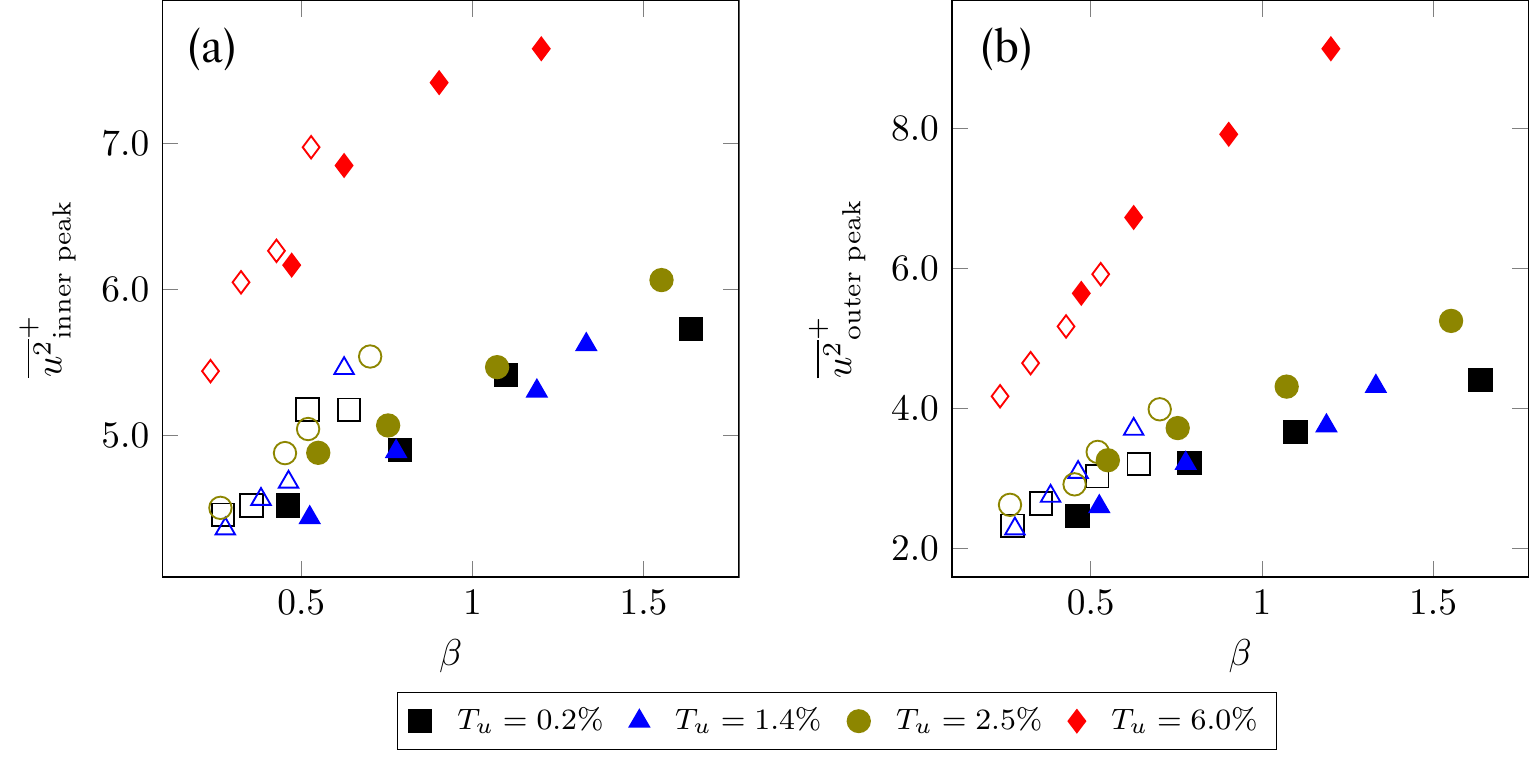}
\end{center}
\caption{Variance (a) inner and (b) outer peak as function of the adverse pressure gradient parameter $\beta$, colours and symbols as in \fref{fig:Cp}, open marks represent AoA of \ang{2}, solid marks represent AoA of \ang{4}.}
\label{fig:peakvariance_beta}
\end{figure}

In \fref{fig:peakvariance_beta}, the magnitudes of the streamwise velocity variance peaks are reported for all investigated conditions (two angles of attack and all freestream turbulence levels) as a function of the pressure-gradient parameter $\beta$. The inner peak is shown in \fref{fig:peakvariance_beta}(a), while the outer peak is reported in \fref{fig:peakvariance_beta}(b). Whereas previous figures focused on individual streamwise locations and the effect of varying the angle of attack, this figure provides a global overview of the variance-peak behavior across all cases.

For both the inner and outer peaks, an increase in peak magnitude is observed with increasing $\beta$, consistent with previous findings for adverse-pressure-gradient turbulent boundary layers \citep{monty2011parametric,harun2013,bobke2017history,lee2017large}. Focusing first on the low freestream turbulence condition (black curve), the growth of both peaks appears approximately linear with $\beta$. This increase can be attributed to the combined effects of the strengthening adverse pressure gradient and the streamwise development of the turbulent boundary layer along the airfoil. Notably, despite these two coupled mechanisms, the data for a fixed value of $T_u$, exhibit a reasonable collapse when plotted as a function of $\beta$, indicating that the pressure-gradient parameter provides a meaningful scaling for the peak growth.

A moderate scatter is observed in the inner-peak data in \fref{fig:peakvariance_beta}(a), which may be partially attributed to spatial filtering effects associated with the finite length of the hot-wire probe. The two lowest freestream turbulence levels ($T_u = 0.2\%$ and $1.4\%$) yield nearly indistinguishable peak magnitudes across the entire $\beta$ range, suggesting a weak sensitivity to moderate freestream turbulence variation in this regime.

For higher freestream turbulence levels, both inner and outer variance peaks increase more rapidly with $\beta$. While the growth remains approximately linear, the slope becomes steeper as $T_u$ increases, with the effect being particularly pronounced for the outer peak. This behavior indicates that the influence of freestream turbulence on the boundary layer intensifies as the adverse pressure gradient strengthens. In other words, at fixed freestream turbulence intensity, increasing $\beta$ amplifies the impact of external turbulent forcing on the boundary layer, leading to progressively larger deviations from the low-$T_u$ cases. These results suggest that adverse pressure gradients enhance the sensitivity of turbulent boundary layers to freestream turbulence.

\begin{figure}[htbp]
\begin{center}
\includegraphics[width=.8\textwidth]{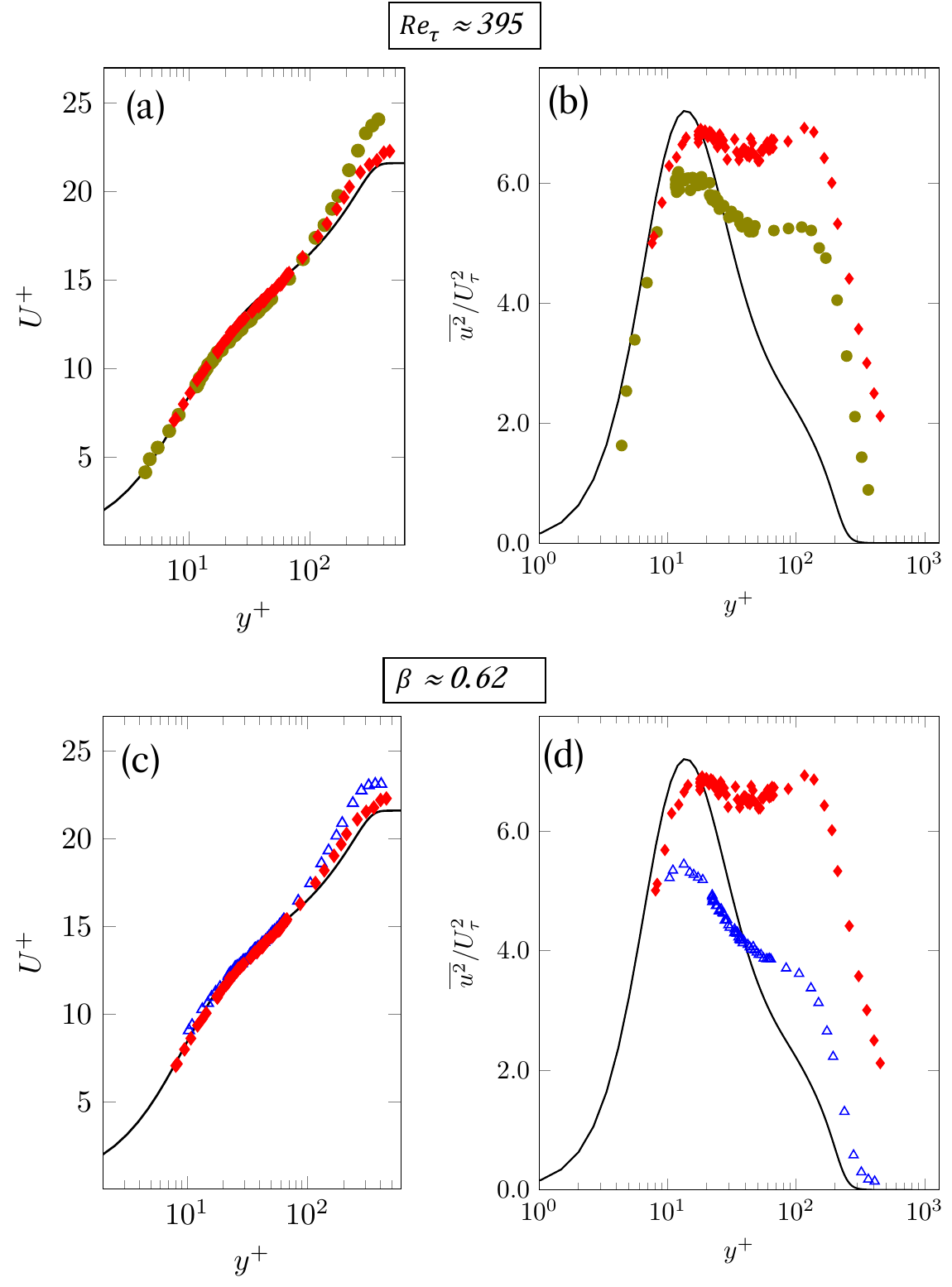}
\end{center}
\caption{Profiles of mean (a,c) and variance (b,d) of the streamwise velocity in wall units; (a,b) effect of $T_u$ on profiles having similar $Re_{\tau} \cong 395$, (c,d) of $T_u$ on profiles having similar $\beta \cong 0.62$.}
\label{fig:profiles_Retau_beta}
\end{figure}

Figure~\fref{fig:profiles_Retau_beta} presents wall-normal profiles of the mean streamwise velocity and the corresponding streamwise velocity variance for cases selected at (a,b) an approximately fixed friction Reynolds number, $Re_{\tau} \approx 395$, and (c,d) an approximately fixed pressure-gradient parameter, $\beta \approx 0.62$. These matching values of $Re_{\tau}$ and $\beta$ are obtained through different combinations of streamwise position, angle of attack, and freestream turbulence intensity.

In \fref{fig:profiles_Retau_beta}(a,b), similar values of $Re_{\tau}$ are achieved for two distinct configurations: $x/c = 0.625$, AoA = $\ang{4}$, $T_u = 2.5\%$ (green solid line) and $x/c = 0.475$, AoA = $\ang{4}$, $T_u = 6\%$ (red solid line). It is important to note that this value of $Re_{\tau}$ is relatively modest, and for canonical ZPG turbulent boundary layers an outer peak in the streamwise velocity variance would not be expected under such conditions. The presence of an outer peak in both cases is therefore attributed to the combined effects of spatial development and adverse pressure gradient. Notably, despite being measured at a more upstream location—where the effect of boundary layer development is weaker—the case with higher freestream turbulence exhibits a more pronounced outer peak than the lower-$T_u$ case measured further downstream. This indicates that freestream turbulence plays a dominant role in amplifying outer-region turbulent activity.

Figures~\fref{fig:profiles_Retau_beta}(c,d) show profiles corresponding to an approximately fixed value of $\beta \approx 0.62$, obtained from two markedly different flow configurations: $x/c = 0.625$, AoA = $\ang{2}$, $T_u = 1.4\%$ (blue open symbols) and $x/c = 0.475$, AoA = $\ang{4}$, $T_u = 6\%$ (red solid line). Although the pressure-gradient parameter is comparable in both cases, the resulting velocity statistics differ substantially. For the highest freestream turbulence condition, the wake region in the mean velocity profile is strongly reduced and the profile approaches a canonical ZPG-like shape. At the same time, a pronounced two-lobed structure emerges in the variance profile, indicating enhanced turbulent activity in both the inner and outer regions of the boundary layer.

Overall, these results demonstrate that, even when either $Re_{\tau}$ or $\beta$ is held approximately constant, freestream turbulence exerts a dominant influence on the turbulent boundary-layer structure. As observed in previous sections, increasing freestream turbulence leads to wake suppression in the mean velocity profile and to a simultaneous enhancement of both inner and outer variance peaks. These findings highlight that matching either the friction Reynolds number or the pressure-gradient parameter alone is insufficient to fully characterize turbulent boundary layers in realistic experimental configurations. In particular, in wind-tunnel experiments involving adverse pressure gradients, freestream turbulence must be considered as a key parameter governing the shape of the velocity profiles and turbulence statistics.

\subsection{Spectral analysis}
\label{sec:spectra}
\begin{figure}[htbp]
\begin{center}
\includegraphics[width=.9\textwidth]{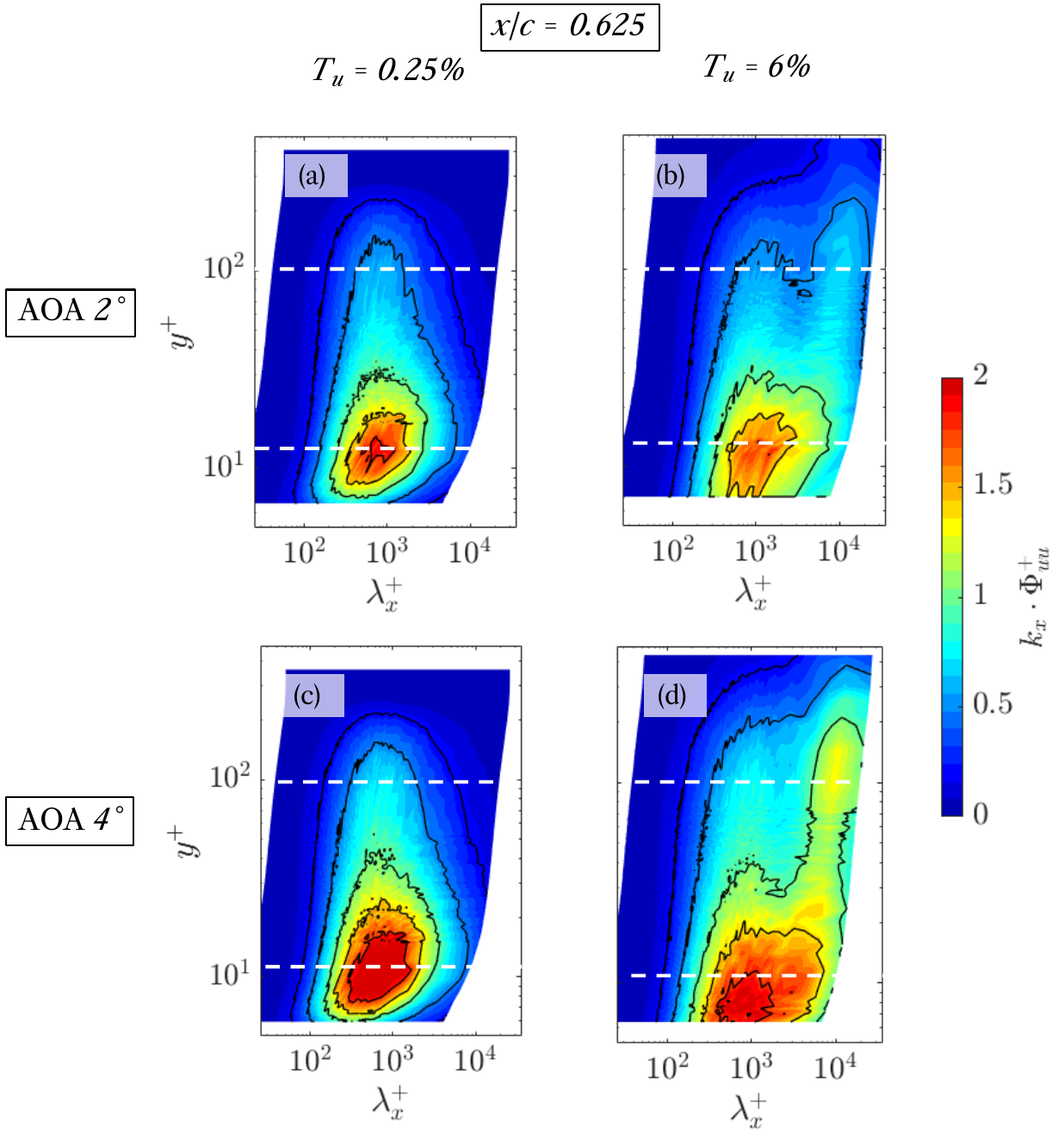}
\end{center}
\caption{Pre-multiplied energy spectra at $x/c=0.625$, comparison of $T_u=0.2\%$ and $T_u=6\%$ for the two angle of attacks.}
\label{fig:spectra_contours}
\end{figure}

\begin{figure}[htbp]
\begin{center}
\includegraphics[width=.85\textwidth]{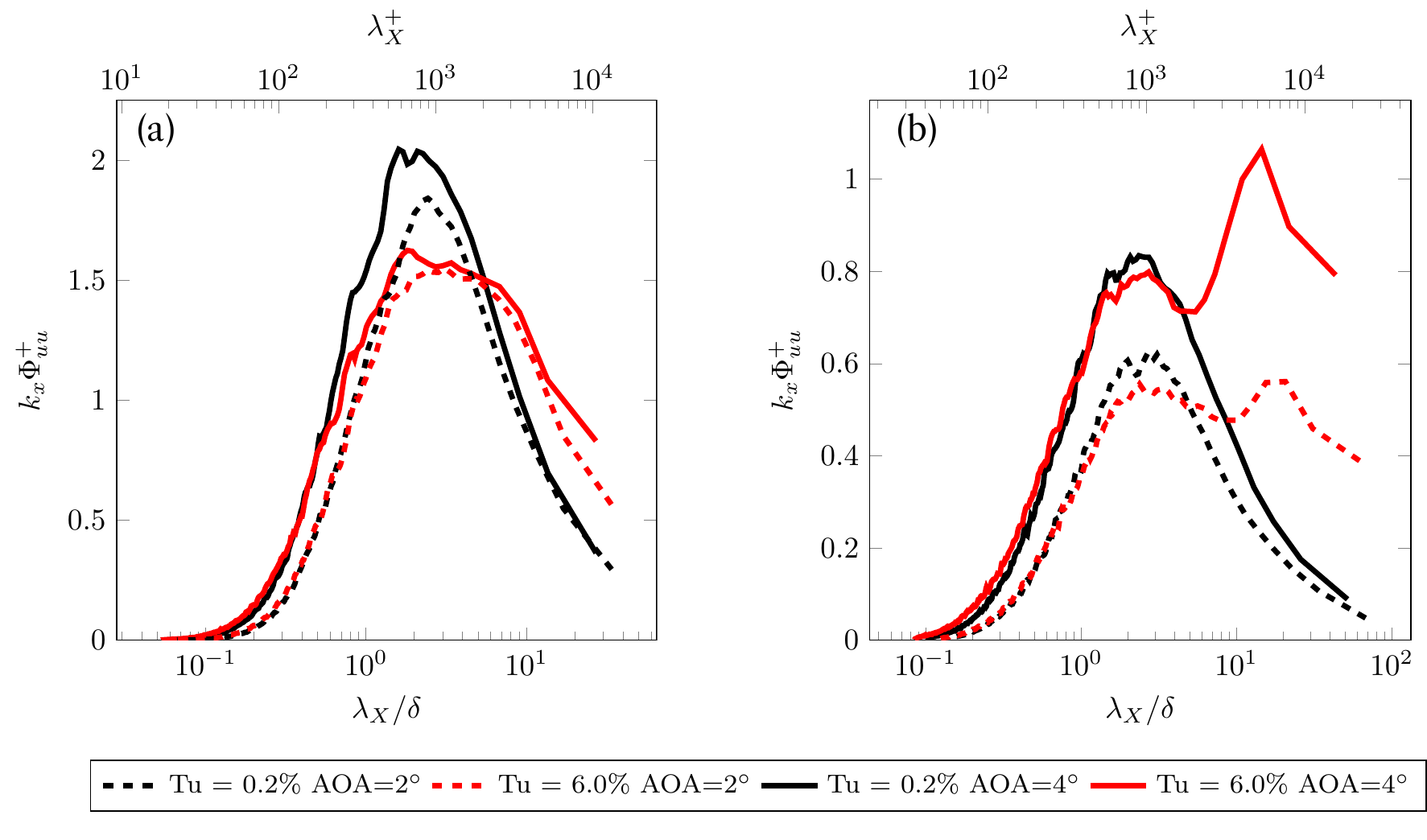}
\end{center}
\caption{Slices of the spectra reported in \fref{fig:spectra_contours} along the dashed white lines, (a) $y^+ \cong 15$ (inner peak) and (b) $y^+ \cong 100$ (outer peak), comparison of $T_u=0.2\%$ and $T_u=6\%$ for the two angles of attacks.}
\label{fig:spectra_slices}
\end{figure}

Figure~\fref{fig:spectra_contours} shows contours of the premultiplied power spectral density of the streamwise velocity fluctuations, $k_x^+ \Phi_{uu}^+$, where $k_x^+$ is the streamwise wavenumber, as a function of wall-normal position and streamwise wavelength $\lambda_x^+$ in wall units. The spectra are obtained from time-resolved measurements using Taylor’s hypothesis. For clarity, only the minimum and maximum freestream turbulence intensities are reported at the most downstream location, $x/c = 0.625$, and the effect of varying the angle of attack is shown. Complementary one-dimensional spectral slices extracted at the near-wall peak location ($y^+ \approx 15$) and in the outer layer ($y^+ \approx 100$) are reported in \fref{fig:spectra_slices} for both freestream turbulence levels and angles of attack.

At low angle of attack and low freestream turbulence intensity (\fref{fig:spectra_contours}a), corresponding to a moderate adverse pressure gradient, a dominant spectral lobe is observed near the wall at $y^+ \approx 15$ and $\lambda_x^+ \approx 10^3$. This feature is associated with the near-wall streaks \citep{schlatter2010} and is also evident in the spectral slice shown in \fref{fig:spectra_slices}(a), where the low-$T_u$ curve exhibits a peak at $\lambda_x^+ \approx 10^3$, corresponding to $\lambda_x/\delta \approx 1$. The energy content in the outer layer ($y^+ \approx 100$) remains relatively modest under these conditions. Increasing the angle of attack while maintaining low freestream turbulence leads to an increase in the magnitude of both the near-wall spectral peak and the energy content in the outer region, reflecting the strengthening influence of the adverse pressure gradient \citep{bobke2017history,lee2017large,pozuelo2023spectra}.

More pronounced modifications arise when introducing high levels of freestream turbulence. At low angle of attack and high freestream turbulence (\fref{fig:spectra_contours}b), an additional spectral lobe becomes clearly visible in the outer region around $y^+ \approx 100-200$ at wavelengths of $\lambda_x^+ \approx 7 \times 10^3$. This wavelength corresponds to very-large-scale motions and is significantly larger than that associated with the near-wall streaks. As shown in the spectral slices (\fref{fig:spectra_slices}b), this outer-layer energy peak corresponds to $\lambda_x/\delta \approx 13$, matching the characteristic wavelength of the freestream turbulence measured outside the boundary layer (see \fref{fig:Tu_characterisation}).

Importantly, the energy associated with these large wavelengths penetrates towards the near-wall region, leading to a broadening of the spectral at $y^+ \approx 15$. The spectral contours extend over wavelengths larger than $\lambda_x^+ \approx 10^3$, indicating that energetic large-scale motions originating from the freestream contribute directly to the enhancement of the near-wall energy. This behavior is consistent with the increase of both inner and outer variance peaks observed in \fref{fig:profiles_Tu_AoA}(d), suggesting that the amplification of turbulent kinetic energy in both regions can be attributed to the penetration of freestream turbulence-associated large scales into the boundary layer.

The penetration of freestream turbulence is further intensified when increasing the angle of attack, and thus the adverse pressure gradient. As shown in \fref{fig:spectra_contours} (d), the outer-layer spectral lobe associated with the freestream turbulence wavelength becomes more energetic compared to the low-angle-of-attack case. This trend is also evident in the outer-layer spectral slices (\fref{fig:spectra_slices}b), where energy at large wavelengths increasingly dominates over shorter scales. In the near-wall region, enhanced contributions from large-scale motions are also observed, although the spectral slices at $y^+ \approx 15$ alone do not fully capture the extent of these changes. This limitation arises because increasing the adverse pressure gradient shifts energetic large-scale motions closer to the wall, pushing a significant portion of the energy below $y^+ \approx 15$.

Overall, these spectral results indicate that the penetration of freestream turbulence-associated very-large-scale motions into the turbulent boundary layer is strongly enhanced in the presence of adverse pressure gradients. The APG promotes increased interaction between the outer and inner regions of the boundary layer, facilitating the transfer of external turbulent energy towards the near-wall region \citep{harun2013,monty2011parametric,tanarro2020effect,pozuelo2023spectra}. This enhanced inner–outer coupling provides a physical explanation for the observed increases in near-wall turbulent kinetic energy, momentum transport, and skin-friction coefficient reported in the previous sections.

\subsection{Scales decomposition}

Based on the spectral analysis presented in the previous section, the increase in turbulent kinetic energy within the boundary layer can be primarily attributed to large-scale freestream turbulence penetrating into the turbulent boundary layer. This penetration affects the mean and variance profiles and, consequently, all higher-order statistics. To further isolate the contribution of different scale ranges, a scale decomposition of the streamwise velocity variance is performed by reconstructing the variance profiles from filtered energy spectra.

Following the approach used by \citet{schlatter2010assessment}, \citet{mathis2011predictive} and \citet{dogan2016interactions}, the fluctuating streamwise velocity signal is decomposed as
\begin{equation}
u^+(t,y) = u_L^+(t,y) + u_S^+(t,y),
\end{equation}
where \(u^+=u/u_\tau\), \(u_L^+\) denotes the large-scale component and \(u_S^+\) denotes the small-scale component.

In the literature, scale separation is commonly performed using a threshold wavelength of $\lambda_x = \delta$ to distinguish between small and large scales, particularly in studies focusing on high Reynolds-number effects \citep{hutchins2007evidence,mathis_2009, dogan2016interactions}. In the present study, a different threshold of $\lambda_x = 8\delta$ is deliberately chosen. This choice is motivated by the spectral results shown in \fref{fig:spectra_slices}, where $\lambda_x = 8\delta$ corresponds to a local minimum in the premultiplied energy spectra. Using this threshold allows the decomposition to capture the majority of the energy associated with the freestream turbulence, including both its spectral peak and the characteristic wavelengths identified in the external flow.
Thus, \(u_L^+\) contains wavelengths \(\lambda_x>8\delta\), associated primarily with the energetic freestream-turbulence scales, whereas \(u_S^+\) contains wavelengths \(\lambda_x<8\delta\).

\begin{figure}[htb]
\begin{center}
\includegraphics[width=1.\textwidth]{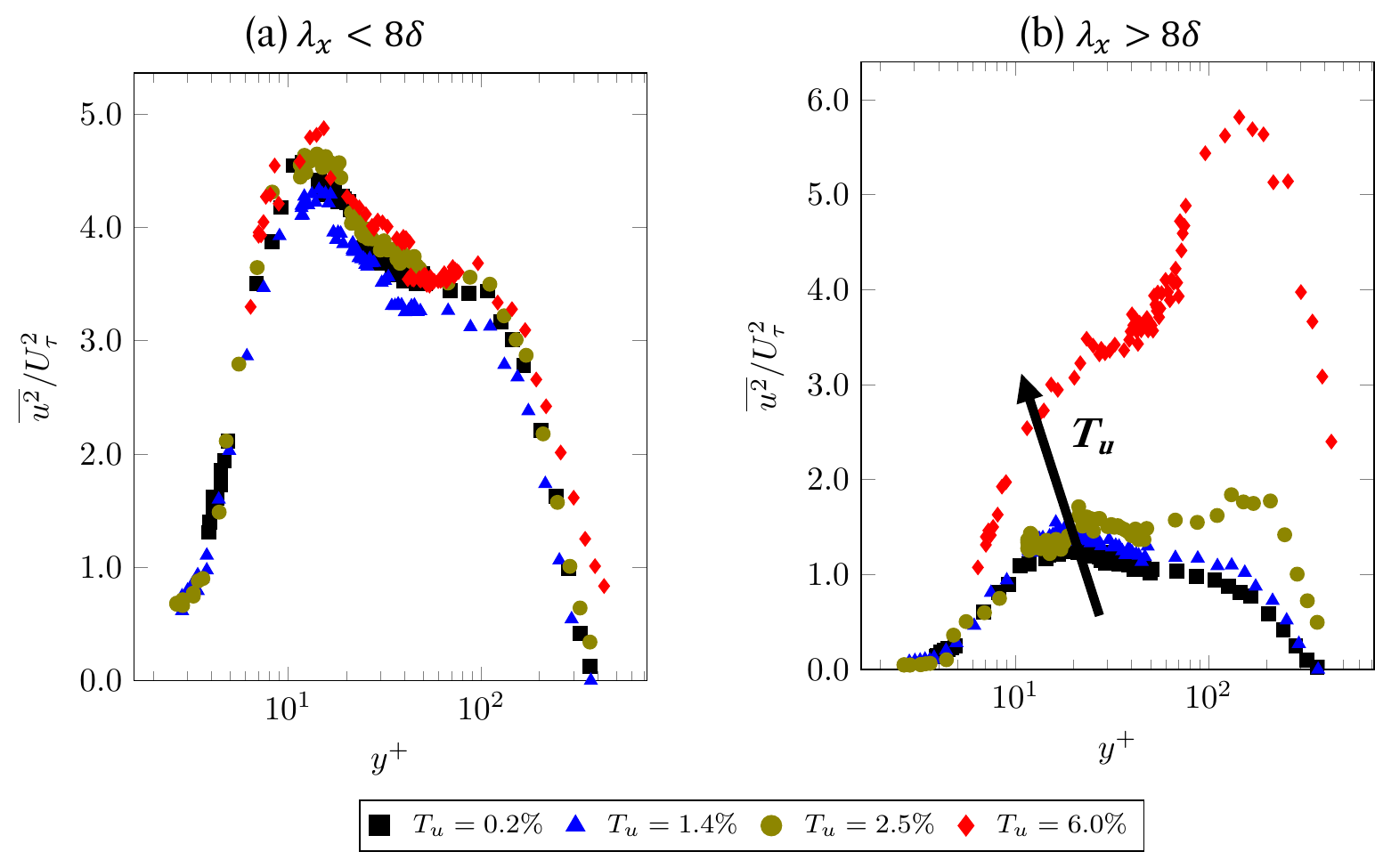}
\end{center}
\caption{Profiles of the variance of the streamwise velocity in wall units at $x/c=0.625$ and $AOA=\ang{4}$ obtained using (a) scales with $\lambda_x<8 \delta$ and (b) $\lambda_x>8 \delta$.}
\label{fig:variance_decomposition}
\end{figure}

Figure~\fref{fig:variance_decomposition} shows the reconstructed variance profiles obtained by filtering wavelengths smaller than $8\delta$ (\fref{fig:variance_decomposition}a) and larger than $8\delta$ (\fref{fig:variance_decomposition}b). For brevity, results are reported only at a fixed streamwise location and at the highest angle of attack, corresponding to the most severe adverse pressure gradient, for all freestream turbulence intensities.

When considering the contribution from scales smaller than the threshold, the variance profiles collapse reasonably well for all freestream turbulence conditions. A weak but systematic increase in variance is observed with increasing $T_u$ throughout the boundary layer, indicating that freestream turbulence has a moderate influence on the small-scale motions. This effect is particularly visible as a slight increase in the near-wall variance peak, suggesting a secondary modulation of the small-scale structures by the external turbulence \citep{dogan2017modelling}.

In contrast, the variance profiles reconstructed from scales larger than $8\delta$ show a strong dependence on the freestream turbulence intensity. The two lowest $T_u$ cases yield nearly identical profiles, whereas increasing $T_u$ leads to a pronounced amplification of the variance, particularly in the outer region around $y^+ \approx 100$–150. The magnitude of the outer peak increases from values of order unity for the lowest freestream turbulence levels to approximately twice that value for $T_u = 2.5\%$, and up to six times for $T_u = 6\%$. Additionally, a clear increase in the near-wall variance peak due to large scales is observed as $T_u$ increases. This result indicates that the enhancement of the near-wall variance peak reported in \fref{fig:profiles_Tu_AoA} can be directly attributed to large-scale structures originating from the freestream and penetrating towards the wall down to $y^+ \approx 15$. The magnitude of this contribution follows a monotonic trend with increasing freestream turbulence intensity.

\begin{figure}[htpb]
\begin{center}
\includegraphics[width=.9\textwidth]{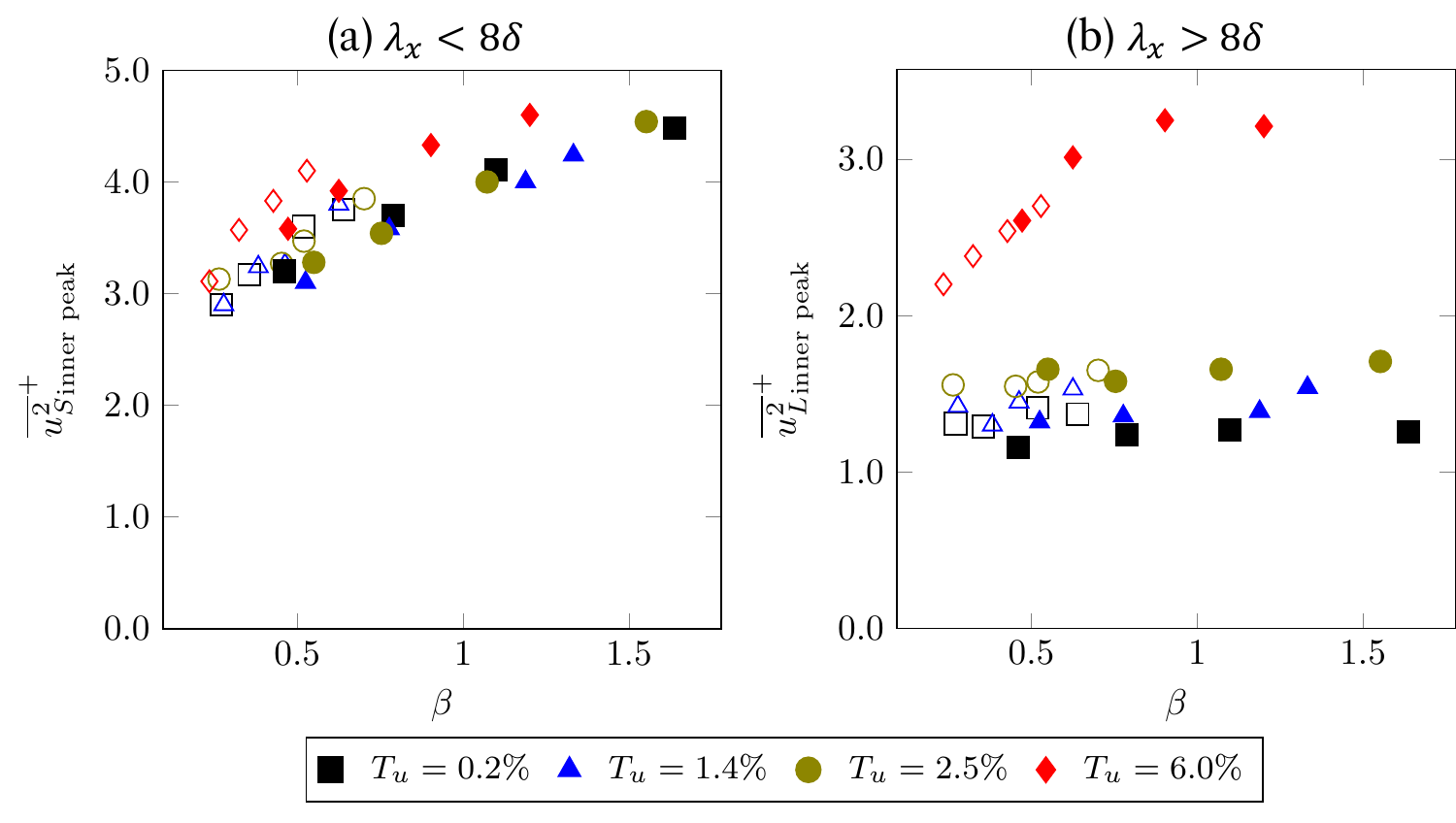}
\end{center}
\caption{Inner peak of the streamwise velocity variance computed using (a) scales with $\lambda_x<8 \delta$ and (b) $\lambda_x>8 \delta$ as function of the pressure gradient parameter $\beta$, open marks represent AoA of \ang{2}, solid marks represent AoA of \ang{4}.}
\label{fig:innerpeak_uLuS_beta}
\end{figure}

Figure~\fref{fig:innerpeak_uLuS_beta} focuses on the near-wall variance peak and summarizes its dependence on the pressure-gradient parameter $\beta$ and freestream turbulence intensity using the complete dataset. In \fref{fig:innerpeak_uLuS_beta}(a), the near-wall peak associated with small-scale motions is reported after filtering out the large scales, while \fref{fig:innerpeak_uLuS_beta}(b) shows the near-wall peak considering only the large-scale contribution associated with the freestream turbulence.

The near-wall peak associated with small scales increases approximately linearly with $\beta$, confirming trends previously reported for adverse-pressure-gradient turbulent boundary layers \citep{monty2011parametric,harun2013,vila2020experimental,pozuelo2023spectra}. This growth can be attributed to the combined effects of the strengthening adverse pressure gradient and the streamwise development of the turbulent boundary layer, which cannot be fully decoupled in the present configuration.
This behavior holds for freestream turbulence intensities up to $T_u = 2.5\%$, whereas a steeper growth is observed for $T_u = 6\%$. This deviation may be partly attributed to uncertainties associated with the filtering procedure or may indicate that, at sufficiently high freestream turbulence levels, the small-scale motions themselves are increasingly influenced by external forcing.

Conversely, the combined effect of freestream turbulence and adverse pressure gradient is more clearly manifested in the large-scale contribution to the near-wall peak. At the lowest freestream turbulence level ($T_u = 0.2\%$), the variation of the large-scale near-wall peak with $\beta$ is weak. As the freestream turbulence intensity increases, the sensitivity of the near-wall peak to $\beta$ becomes progressively stronger. While the $T_u = 1.4\%$ case remains close to the lowest-$T_u$ condition, a clear deviation emerges at $T_u = 2.5\%$, and for $T_u = 6\%$ the growth rate with $\beta$ is substantially enhanced. At the highest freestream turbulence level, the growth appears to saturate for $\beta \gtrsim 0.5$, suggesting a possible limit to the penetration of freestream turbulence into the near-wall region. Further work is required to clarify the physical origin of this saturation, and to determine whether it reflects a genuine flow mechanism or is influenced by Reynolds-number, measurement, or filtering limitations.

These results indicate that adverse pressure gradients facilitate the penetration of large-scale freestream turbulence into the turbulent boundary layer. For a given freestream turbulence intensity—thus for fixed external length scales and energy—an increase in $\beta$ leads to a stronger deviation from the low-$T_u$ behavior, with a more pronounced impact on the near-wall variance peak. This suggests that adverse pressure gradients, by enhancing inner–outer layer interactions, promote the influence of large-scale external turbulence on near-wall dynamics.

\subsection{Skewness decomposition and scale interactions}
\label{sec:skewness_decomposition}

The spectral and variance decompositions discussed above show that large-scale motions associated with the freestream turbulence penetrate the boundary layer and contribute to both the outer and near-wall turbulence intensity. However, an increase in near-wall variance may arise from two distinct mechanisms: a direct superposition of large-scale velocity fluctuations in the near-wall region, or a modulation of the small-scale near-wall cycle by the large-scale motions. 

Scale interactions in wall-bounded turbulence have often been analysed in terms of the influence of energetic large-scale motions on the near-wall small scales. Early evidence of coupling between low- and high-frequency motions was reported by \citet{bandyopadhyay1984coupling}. Subsequent studies showed that large scales leave a near-wall footprint and modulate the amplitude of the small-scale fluctuations \citep{hutchins2007large,mathis_2009}. This modulation is commonly quantified through the correlation between the large-scale velocity fluctuation and the envelope of the small-scale signal \citep{mathis_2009,ganapathisubramani2012amplitude}. Related approaches have also been used to develop predictive inner-outer models of near-wall turbulence based on large-scale information \citep{marusic2010predictive,mathis2011predictive}. More recently, \citet{jacobi2013phase} and \citet{duvvuri2015triadic} showed that externally imposed large-scale disturbances can modify the phase relationship between large and small scales, which can be diagnosed using both amplitude-modulation coefficients and skewness-based quantities \citep{chung2010large,ganapathisubramani2012amplitude}.

To highligths a modulation of the small-scale near-wall cycle by the large-scale motions, following \citet{dogan2016interactions} and \citet{mathis2011relationship}, the skewness of the streamwise velocity fluctuations is decomposed into contributions associated with large and small scales

\begin{equation}
S_u =
\overline{\overline{u_L^{+3}}}
+
3\overline{\overline{u_L^{+2}u_S^+}}
+
3\overline{\overline{u_L^+u_S^{+2}}}
+
\overline{\overline{u_S^{+3}}},
\label{eq:skewness_decomposition}
\end{equation}
where the double overbar denotes normalization by \(\overline{u^{+2}}^{3/2}\).

\begin{figure}[htbp]
\begin{center}
\includegraphics[width=1\textwidth]{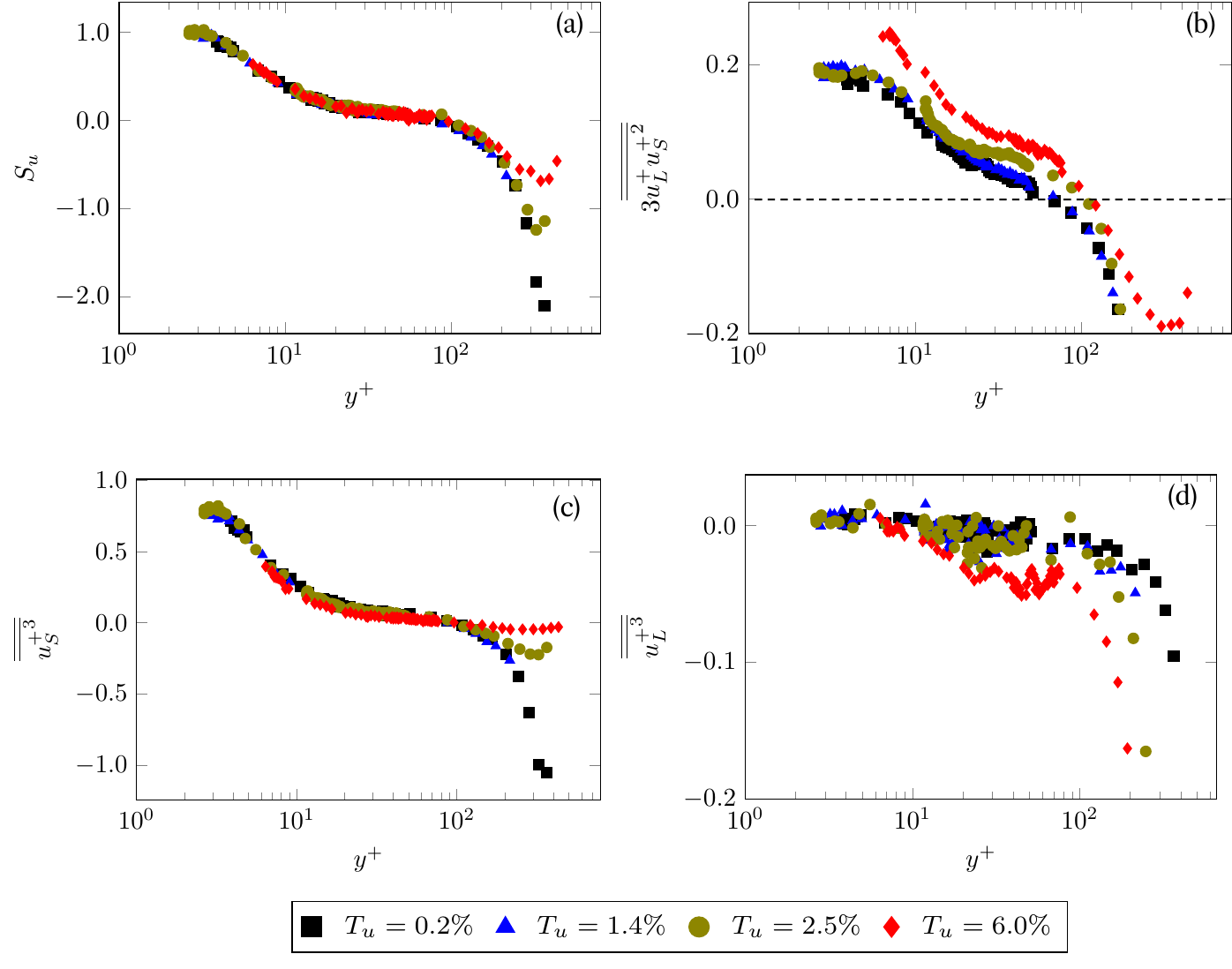}
\end{center}
\caption{Skewness decomposition of the streamwise velocity fluctuations at $x/c=0.625$ and $\alpha=\ang{4}$ for different freestream turbulence intensities. Subfigure (a) shows the total skewness $S_u$, while subfigures (b-d) show the scale-decomposed contributions defined in \eqref{eq:skewness_decomposition}: (b) the modulation term $3\overline{\overline{u_L^+u_S^{+2}}}$, (c) the small-scale contribution $\overline{\overline{u_S^{+3}}}$, and (d) the large scale term $\overline{\overline{u_L^{+3}}}$. Colours and symbols are as in \fref{fig:Cp}.}
\label{fig:skew_dec}
\end{figure}

Figure~\ref{fig:skew_dec} reports the total skewness and the scale-decomposed contributions defined in \eqref{eq:skewness_decomposition} at the most downstream location, $x/c=0.625$, and for $\alpha=\ang{4}$. Only this condition is shown for brevity, since it corresponds to the strongest APG and therefore to the case where the interaction between the external turbulence and the boundary layer is expected to be most pronounced.
The second term on the right hand side, \(3\overline{u_L^{+2}u_S^+}\), is close to zero in agreement with \citet{dogan2016interactions} and to contribute only weakly to the total skewness and is therefore not reported here.

The total skewness profiles, shown in \fref{fig:skew_dec}(a), exhibit a good collapse over most of the boundary layer for all freestream turbulence levels. Differences are mainly observed in the outer region, for $y^+ \gtrsim 150$, where the highest-\(T_u\) case shows lower values of skewness. This reduction is consistent with the decrease of outer-region intermittency reported in turbulent boundary layers subjected to freestream turbulence \citep{blair1983influence2,sharp2009effects,dogan2016interactions}. 

The term \(3\overline{u_L^+u_S^{+2}}\), reported in \fref{fig:skew_dec}(b), is the most relevant term for assessing modulation effects. According to \citet{mathis2011relationship, mathis_2009} is closely related to the correlation between the large-scale velocity fluctuation and the filtered envelope of the small-scale fluctuations.
It shows the clearest dependence on freestream turbulence intensity. At a fixed wall-normal location, this contribution increases systematically with \(T_u\), indicating that the large-scale motions associated with the freestream turbulence increasingly modulate the small-scale fluctuations. The zero crossing is also shifted towards larger wall-normal distances as \(T_u\) increases, showing that positive large-scale motions and enhanced small-scale activity remain in phase over a larger portion of the boundary layer. This behaviour supports the interpretation that the large scales introduced by the freestream turbulence do not simply superpose onto the near-wall signal, but actively modulate the near-wall small-scale motions.

The small-scale contribution, \(\overline{u_S^{+3}}\), shown in \fref{fig:skew_dec}(c), collapses reasonably well for the different freestream turbulence levels over the inner part of the boundary layer. Deviations are mainly observed for \(y^+ \gtrsim 100\), where the separation between small and large scales becomes less sharp and where the influence of the energetic outer motions becomes more important. This collapse indicates that, once the large-scale contribution is removed using the present cut-off at \(\lambda_x=8\delta\), the intrinsic skewness of the small-scale component is only weakly modified by the freestream turbulence.
Finally, the term $\overline{\overline{u_L^{+3}}}$, reported in \fref{fig:skew_dec}(d), remains close to zero throughout most of the boundary layer. This behaviour is consistent with the observations of \citet{dogan2016interactions}, who found this contribution to be small compared with the other terms. 

Overall, the decomposition shows that the increase of near-wall turbulent activity under elevated freestream turbulence is not only due to the direct penetration of large-scale energy into the boundary layer. The systematic growth of \(3\overline{u_L^+u_S^{+2}}\) demonstrates that the external large scales also modulate the near-wall small-scale motions.

\section{Concluding remarks}
\label{sec:conclusions}

The present work investigated the combined effect of freestream turbulence and adverse pressure gradient on turbulent boundary layers developing over a NACA 0015 airfoil. Measurements were performed at fixed chord Reynolds number, for two angles of attack and four freestream turbulence levels, over the chordwise range $x/c=0.400$-$0.625$. This configuration allowed the response of a spatially developing APG TBL to be assessed under controlled levels of external turbulence.

Increasing the freestream turbulence level systematically modified the mean-flow development. The boundary layer became thicker, the shape factor decreased and the wake region of the mean velocity profile was partially suppressed. At the same time, the skin-friction coefficient increased and approached values closer to the canonical ZPG, despite the presence of an adverse pressure gradient. These trends show that FST counteracts several mean-flow effects usually associated with APG TBLs, namely wake growth, shape-factor increase and skin-friction reduction.

\begin{figure}[htbp]
\begin{center}
\includegraphics[width=.7\textwidth]{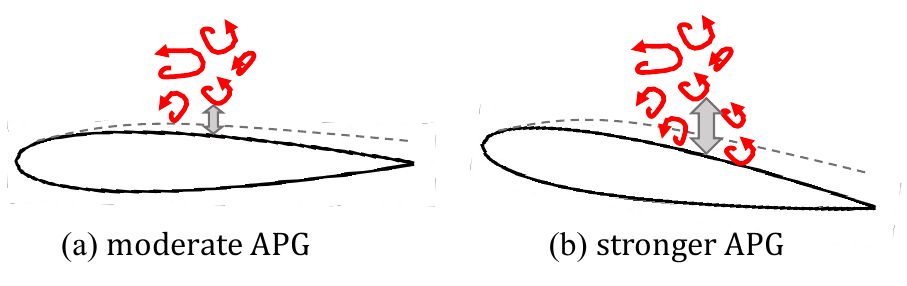}
\end{center}
\caption{Schematic of the interaction between freestream turbulence and an APG airfoil boundary layer. Increasing the APG promotes the penetration of energetic freestream motions into the boundary layer and strengthens their effect on the near-wall region.}
\label{fig:schematic_conclusions}
\end{figure}

The imposed freestream turbulence also produced substantial changes in the streamwise velocity variance. Both the near-wall and outer peaks increased with FST intensity. Spectral analysis showed that this amplification is associated with energetic large-scale motions introduced by the freestream turbulence, with characteristic wavelengths of order $\lambda_x/\delta \approx 13$. These motions penetrate into the boundary layer and contribute not only to the outer-layer energy but also to the near-wall variance. The scale decomposition confirmed that the large-scale contribution to the near-wall variance becomes increasingly important as both \(T_u\) and \(\beta\) increase, indicating that the adverse pressure gradient enhances the influence of freestream-turbulence scales on the near-wall region.
The skewness decomposition further showed that the externally introduced large scales do not simply superpose onto the near-wall signal. The modulation term increased systematically with freestream turbulence intensity, indicating enhanced coupling between large-scale motions and near-wall small-scale activity. 

Overall, the current results indicate that stronger APG conditions promote the penetration of energetic freestream motions and strengthen their interaction with the near-wall region (\fref{fig:schematic_conclusions}). In this sense, the APG boundary layer appears more receptive to large-scale freestream turbulence than a comparable ZPG boundary layer.
More speculatively, the combined action of APG and FST may allow some features usually associated with higher-Reynolds-number wall turbulence to emerge even at the relatively modest friction Reynolds numbers investigated here, including enhanced outer-layer energy, increased inner--outer interaction and modulation of near-wall fluctuations.

These findings have direct implications for realistic aerodynamic configurations, such as airfoils and drones operating in urban or atmospheric turbulence. In such conditions, the dominant energetic structures affecting the near-wall region may be imposed by the external turbulence and amplified by the adverse pressure gradient. Therefore, FST should be treated as a governing parameter in APG airfoil boundary layers, particularly when interpreting skin friction, separation, turbulence statistics and scale interactions. 
The results also suggest that, even at moderate Reynolds numbers, flow-control strategies designed for canonical boundary layers may need to account for the enhanced role of large-scale motions under combined APG and FST conditions.

\newpage\clearpage

\renewcommand{\thefigure}{A.\arabic{figure}}
\setcounter{figure}{0}

\section*{Appendix A: determination of the boundary layer thickness}
\label{sec:appendixA}

The determination of the boundary-layer thickness is particularly delicate in the present measurements because the velocity profiles are affected simultaneously by adverse pressure gradient and freestream turbulence. The boundary-layer thickness is required for the computation of integral quantities, such as the displacement thickness, momentum thickness and shape factor, and is also used to normalize the spectra and to define the scale-separation threshold adopted in the present analysis.

\begin{figure}[htbp]
\begin{center}
\includegraphics[width=.7\textwidth]{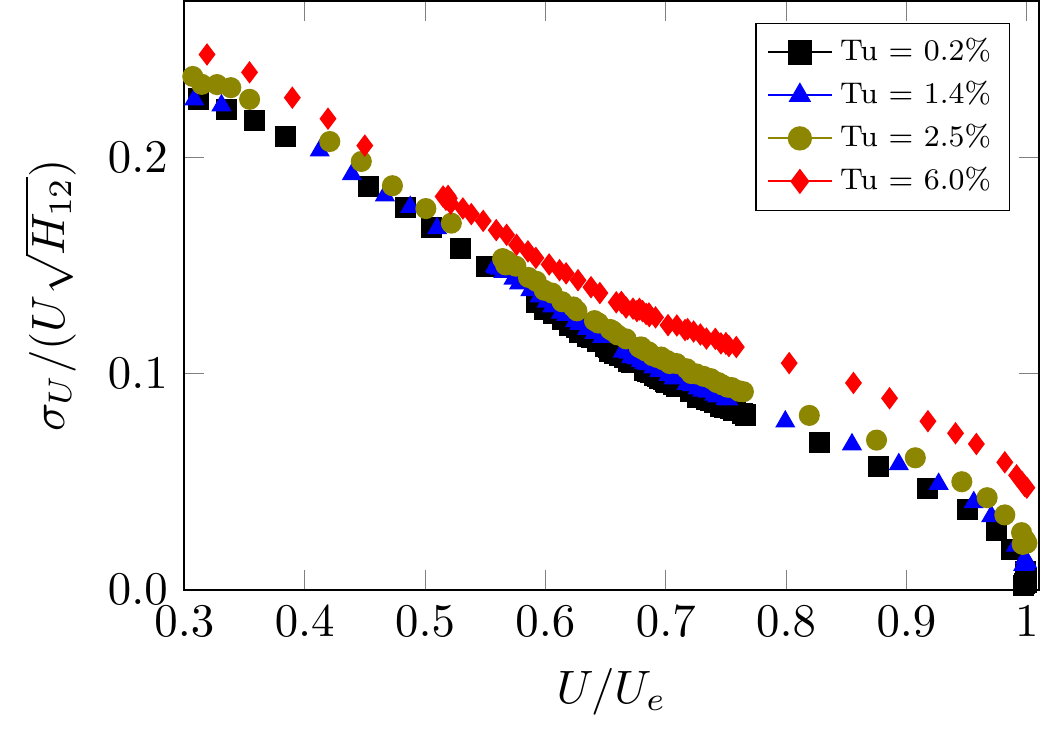}
\end{center}
\caption{Diagnostic function at $x/c=0.4$, $AOA=\ang{2}$ at different $Tu$.}
\label{fig:diagnostic}
\end{figure}

A commonly used approach for pressure-gradient turbulent boundary layers is the diagnostic-plot method proposed by \citet{vinuesa_determining_2016}. In this method, the streamwise turbulence intensity is plotted as a function of the normalized mean velocity, and the boundary-layer edge is identified from the point corresponding to \(U/U_e=0.99\). For pressure-gradient turbulent boundary layers, \citet{vinuesa_determining_2016} showed that this condition corresponds approximately to
\begin{equation}
\frac{u'}{U\sqrt{H_{12}}} \simeq 0.02,
\end{equation}
where \(H_{12}=\delta^\ast/\theta\). Since \(U_e\), \(H_{12}\) and \(\delta_{99}\) are not known a priori, the method is applied iteratively: an initial estimate of \(U_e\) is used to construct the diagnostic plot, \(\delta_{99}\) is then identified from the above threshold, and the procedure is repeated until convergence.

This approach is robust for pressure-gradient turbulent boundary layers when the turbulence intensity decreases towards the freestream and the boundary-layer edge is clearly identified in the diagnostic plot. In the present measurements, however, the imposed freestream turbulence produces a finite turbulence level outside the boundary layer. As a consequence, the diagnostic function does not approach a near-zero freestream value, and the threshold \({u'}/({U\sqrt{H_{12}}})\simeq0.02\) becomes sensitive to the imposed FST level. Already for moderate values of \(T_u\), the diagnostic plot no longer provides a unique and robust identification of \(\delta_{99}\), as illustrated in \fref{fig:diagnostic}. 

For this reason, the boundary-layer thickness used in the present work is obtained from the fitting procedure of \citet{rodriguez-lopez_robust_2015}. In this approach, \(\delta\) is treated as an open fitting parameter in the mean-velocity-profile fit, together with the friction velocity. This avoids imposing a fixed threshold at the boundary-layer edge and provides a more stable estimate of \(\delta\) for profiles affected by both APG and FST. Once \(\delta\) is obtained from the fit, the displacement and momentum thicknesses are computed by direct integration of the measured mean velocity profiles, and the shape factor is evaluated as
\begin{equation}
H = \frac{\delta^\ast}{\theta}.
\end{equation}

\section*{Appendix B: Streamwise development of the mean and variance profiles}
\label{sec:appendixB}

\renewcommand{\thefigure}{B.\arabic{figure}}
\setcounter{figure}{0}

\begin{figure}[htbp]
\begin{center}
\includegraphics[width=.9\textwidth]{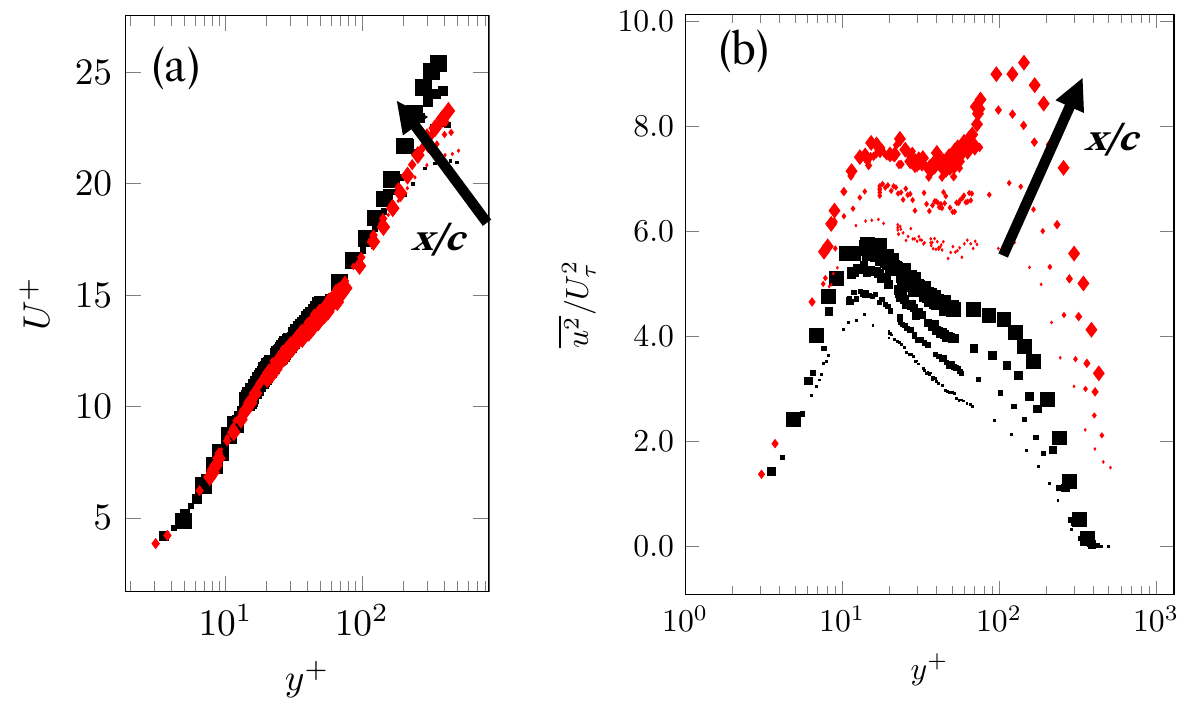}
\end{center}
\caption{Streamwise evolution of the wall-normal profiles of (a) mean streamwise velocity and (b) streamwise velocity variance in wall units at $\alpha=\ang{4}$. The lowest and highest freestream turbulence levels are reported. Larger symbols indicate more downstream locations.}
\label{fig:profiles_streamwise_AoA4}
\end{figure}

Figure~\ref{fig:profiles_streamwise_AoA4} shows the streamwise evolution of the mean streamwise velocity and variance profiles at $\alpha=\ang{4}$ for the lowest and highest freestream turbulence levels. This angle of attack is selected because it corresponds to the strongest adverse-pressure-gradient condition investigated in the present study. The larger symbols indicate progressively downstream measurement locations.

For the lowest freestream turbulence level, the mean velocity profiles show a progressively stronger wake component as the flow develops downstream. This behaviour is consistent with the increase of the adverse pressure gradient along the airfoil and with the expected response of APG turbulent boundary layers. The variance profiles also show an increase of both the near-wall and outer-layer peaks with downstream distance. This trend reflects the combined effect of the spatial development of the turbulent boundary layer and the increasing adverse pressure gradient, consistent with previous observations of developing TBLs and APG-induced outer-layer energization \citep{vincenti2013streamwise,marusic2015evolution,monty2011parametric,harun2013,bobke2017history,lee2017large}.

At the highest freestream turbulence level, the mean profiles remain closer to the canonical ZPG behaviour, indicating the wake-suppressing effect of FST discussed in the main text. Nevertheless, the variance profiles show a strong amplification of turbulent activity, especially in the outer region, as the flow develops downstream. The most downstream station, $x/c=0.625$, therefore represents the condition where the combined effects of APG, spatial development and freestream turbulence are most pronounced.

\newpage\clearpage

\renewcommand\refname{References}

\bibliography{biblio.bib}

@article{schlatter2010,
  title={Assessment of direct numerical simulation data of turbulent boundary layers},
  author={Schlatter, P. and Orlu, R.},
  journal={Journal of Fluid Mechanics},
  volume={659},
  pages={116--126},
  year={2010},
  publisher={Cambridge University Press}
}

@article{FERNHOLZ1996245,
title = {The incompressible zero-pressure-gradient turbulent boundary layer: An assessment of the data},
journal = {Progress in Aerospace Sciences},
volume = {32},
number = {4},
pages = {245-311},
year = {1996},
issn = {0376-0421},
doi = {https://doi.org/10.1016/0376-0421(95)00007-0},
url = {https://www.sciencedirect.com/science/article/pii/0376042195000070},
author = {H.H. Fernholz and P.J. Finleyt}
}

@article{mathis_2009, title={Large-scale amplitude modulation of the small-scale structures in turbulent boundary layers}, volume={628}, DOI={10.1017/S0022112009006946}, journal={Journal of Fluid Mechanics}, publisher={Cambridge University Press}, author={Mathis, Romain and Hutchins, Nicholas and Marusic, Ivan}, year={2009}, pages={311–337}}

@article{hutchins2007evidence,
  title={Evidence of very long meandering features in the logarithmic region of turbulent boundary layers},
  author={Hutchins, Nicholas and Marusic, Ivan},
  journal={Journal of Fluid Mechanics},
  volume={579},
  pages={1--28},
  year={2007},
  publisher={Cambridge University Press}
}

@article{hancock1989turbulence,
  title={Turbulence structure of a boundary layer beneath a turbulent free stream},
  author={Hancock, PE and Bradshaw, P},
  journal={Journal of Fluid Mechanics},
  volume={205},
  pages={45--76},
  year={1989},
  publisher={Cambridge University Press}
}

@article{castro1984effects,
  author  = {Castro, I. P.},
  title   = {Effects of Free Stream Turbulence on Low Reynolds Number Boundary Layers},
  journal = {Journal of Fluids Engineering},
  year    = {1984},
  volume  = {106},
  number  = {3},
  pages   = {298--306}
}

@article{isaza2014grid,
  title={On grid-generated turbulence in the near-and far field regions},
  author={Isaza, Juan C and Salazar, Ricardo and Warhaft, Zellman},
  journal={Journal of Fluid Mechanics},
  volume={753},
  pages={402--426},
  year={2014},
  publisher={Cambridge University Press}
}

@article{kozul2020response,
  title={Response of the temporal turbulent boundary layer to decaying free-stream turbulence},
  author={Kozul, Melissa and Hearst, R Jason and Monty, Jason P and Ganapathisubramani, Bharathram and Chung, Daniel},
  journal={Journal of Fluid Mechanics},
  volume={896},
  pages={A11},
  year={2020},
  publisher={Cambridge University Press}
}

@article{hearst2018robust,
  title={Robust features of a turbulent boundary layer subjected to high-intensity free-stream turbulence},
  author={Hearst, R Jason and Dogan, Eda and Ganapathisubramani, Bharathram},
  journal={Journal of Fluid Mechanics},
  volume={851},
  pages={416--435},
  year={2018},
  publisher={Cambridge University Press}
}

@article{jooss2021spatial,
  title={Spatial development of a turbulent boundary layer subjected to freestream turbulence},
  author={Jooss, Yannick and Li, Leon and Bracchi, Tania and Hearst, R Jason},
  journal={Journal of Fluid Mechanics},
  volume={911},
  pages={A4},
  year={2021},
  publisher={Cambridge University Press}
}

@article{hancock1983effect,
  author  = {Hancock, P. E. and Bradshaw, P.},
  title   = {The Effect of Free-Stream Turbulence on Turbulent Boundary Layers},
  journal = {Journal of Fluids Engineering},
  year    = {1983},
  volume  = {105},
  number  = {3},
  pages   = {284--289}
}

@article{blair1983influence2,
  author  = {Blair, M. F.},
  title   = {Influence of Free-Stream Turbulence on Turbulent Boundary Layer Heat Transfer and Mean Profile Development: Part II---Analysis of Results},
  journal = {Journal of Heat Transfer},
  year    = {1983},
  volume  = {105},
  number  = {1},
  pages   = {41--47},
  doi     = {10.1115/1.3245557}
}

@article{blair1983influence1,
  author  = {Blair, M. F.},
  title   = {Influence of Free-Stream Turbulence on Turbulent Boundary Layer Heat Transfer and Mean Profile Development: Part I---Experimental Data},
  journal = {Journal of Heat Transfer},
  year    = {1983},
  volume  = {105},
  number  = {1},
  pages   = {33--40},
  doi     = {10.1115/1.3245555}
}

@article{thole1996high,
  author  = {Thole, K. A. and Bogard, D. G.},
  title   = {High Freestream Turbulence Effects on Turbulent Boundary Layers},
  journal = {Journal of Fluids Engineering},
  year    = {1996},
  volume  = {118},
  number  = {2},
  pages   = {276--284}
}

@article{sharp2009effects,
  title={Effects of large-scale free stream turbulence on a turbulent boundary layer},
  author={Sharp, Nicole Susanne and Neuscamman, S and Warhaft, Z},
  journal={Physics of Fluids},
  volume={21},
  number={9},
  year={2009},
  publisher={AIP Publishing}
}

@article{dogan2016interactions,
  title={Interactions of large-scale free-stream turbulence with turbulent boundary layers},
  author={Dogan, Eda and Hanson, Ronald E and Ganapathisubramani, Bharathram},
  journal={Journal of Fluid Mechanics},
  volume={802},
  pages={79--107},
  year={2016},
  publisher={Cambridge University Press}
}

@article{dogan2017modelling,
  title={Modelling high Reynolds number wall--turbulence interactions in laboratory experiments using large-scale free-stream turbulence},
  author={Dogan, Eda and Hearst, R Jason and Ganapathisubramani, Bharathram},
  journal={Philosophical Transactions of the Royal Society A: Mathematical, Physical and Engineering Sciences},
  volume={375},
  number={2089},
  pages={20160091},
  year={2017},
  publisher={The Royal Society Publishing}
}

@article{esteban2017skin,
  title={Skin-friction measurements in a turbulent boundary layer under the influence of free-stream turbulence},
  author={Esteban, Luis Blay and Dogan, Eda and Rodr{\'\i}guez-L{\'o}pez, Eduardo and Ganapathisubramani, Bharathram},
  journal={Experiments in Fluids},
  volume={58},
  pages={1--7},
  year={2017},
  publisher={Springer}
}

@article{vincenti2013streamwise,
  title={Streamwise velocity statistics in turbulent boundary layers that spatially develop to high Reynolds number},
  author={Vincenti, P and Klewicki, J and Morrill-Winter, C and White, CM and Wosnik, M},
  journal={Experiments in fluids},
  volume={54},
  pages={1--13},
  year={2013},
  publisher={Springer}
}

@article{vinuesa_determining_2016,
	title = {On determining characteristic length scales in pressure-gradient turbulent boundary layers},
	volume = {28},
	issn = {1070-6631, 1089-7666},
	url = {https://pubs.aip.org/pof/article/28/5/055101/1079422/On-determining-characteristic-length-scales-in},
	doi = {10.1063/1.4947532},
	language = {en},
	number = {5},
	urldate = {2023-09-13},
	journal = {Physics of Fluids},
	author = {Vinuesa, R. and Bobke, A. and Örlü, R. and Schlatter, P.},
	month = may,
	year = {2016},
	pages = {055101},
}

@article{rodriguez-lopez_robust_2015,
	title = {A robust post-processing method to determine skin friction in turbulent boundary layers from the velocity profile},
	volume = {56},
	issn = {0723-4864, 1432-1114},
	url = {http://link.springer.com/10.1007/s00348-015-1935-5},
	doi = {10.1007/s00348-015-1935-5},
	language = {en},
	number = {4},
	urldate = {2024-01-11},
	journal = {Experiments in Fluids},
	author = {Rodríguez-López, Eduardo and Bruce, Paul J. K. and Buxton, Oliver R. H.},
	month = apr,
	year = {2015},
	pages = {68},
}

@article{marusic2015evolution,
  title={Evolution of zero-pressure-gradient boundary layers from different tripping conditions},
  author={Marusic, I and Chauhan, KA and Kulandaivelu, V and Hutchins, N},
  journal={Journal of Fluid Mechanics},
  volume={783},
  pages={379--411},
  year={2015},
  publisher={Cambridge University Press}
}

@article{raushan2018grid,
  title={Grid generated turbulence under the rigid boundary influence},
  author={Raushan, Pankaj Kumar and Singh, Santosh Kumar and Debnath, Koustuv},
  journal={Journal of Wind Engineering and Industrial Aerodynamics},
  volume={182},
  pages={252--261},
  year={2018},
  publisher={Elsevier}
}

@article{you2019conditional,
  title={Conditional statistics and flow structures in turbulent boundary layers buffeted by free-stream disturbances},
  author={You, Jiho and Zaki, Tamer A},
  journal={Journal of fluid Mechanics},
  volume={866},
  pages={526--566},
  year={2019},
  publisher={Cambridge University Press}
}

@article{monty2011parametric,
  title={A parametric study of adverse pressure gradient turbulent boundary layers},
  author={Monty, Jason P and Harun, Zambri and Marusic, Ivan},
  journal={International Journal of Heat and Fluid Flow},
  volume={32},
  number={3},
  pages={575--585},
  year={2011},
  publisher={Elsevier}
}

@article{vila2020experimental,
  title={Experimental realisation of near-equilibrium adverse-pressure-gradient turbulent boundary layers},
  author={Vila, C Sanmiguel and Vinuesa, Ricardo and Discetti, Stefano and Ianiro, Andrea and Schlatter, Philipp and {\"O}rl{\"u}, Ramis},
  journal={Experimental Thermal and Fluid Science},
  volume={112},
  pages={109975},
  year={2020},
  publisher={Elsevier}
}

@article{bobke2017history,
  title={History effects and near equilibrium in adverse-pressure-gradient turbulent boundary layers},
  author={Bobke, Alexandra and Vinuesa, Ricardo and {\"O}rl{\"u}, Ramis and Schlatter, Philipp},
  journal={Journal of Fluid Mechanics},
  volume={820},
  pages={667--692},
  year={2017},
  publisher={Cambridge University Press}
}

@article{vinuesa2017revisiting,
  title={Revisiting history effects in adverse-pressure-gradient turbulent boundary layers},
  author={Vinuesa, Ricardo and {\"O}rl{\"u}, Ramis and Sanmiguel Vila, Carlos and Ianiro, Andrea and Discetti, Stefano and Schlatter, Philipp},
  journal={Flow, turbulence and combustion},
  volume={99},
  pages={565--587},
  year={2017},
  publisher={Springer}
}

@article{tanarro2020effect,
  title={Effect of adverse pressure gradients on turbulent wing boundary layers},
  author={Tanarro, {\'A}lvaro and Vinuesa, Ricardo and Schlatter, Philipp},
  journal={Journal of Fluid Mechanics},
  volume={883},
  pages={A8},
  year={2020},
  publisher={Cambridge University Press}
}

@article{volino2020non,
	title = {Non-equilibrium development in turbulent boundary layers with changing pressure gradients},
	volume = {897},
	issn = {0022-1120, 1469-7645},
	url = {https://www.cambridge.org/core/product/identifier/S0022112020003195/type/journal_article},
	doi = {10.1017/jfm.2020.319},
	language = {en},
	urldate = {2023-11-20},
	journal = {Journal of Fluid Mechanics},
	author = {Volino, Ralph J.},
	month = aug,
	year = {2020},
	pages = {A2},
}

@article{harun2013, title={Pressure gradient effects on the large-scale structure of turbulent boundary layers}, volume={715}, DOI={10.1017/jfm.2012.531}, journal={Journal of Fluid Mechanics}, author={Harun, Zambri and Monty, Jason P. and Mathis, Romain and Marusic, Ivan}, year={2013}, pages={477–498}}

@article{pozuelo2023spectra,
	title = {Spectra of near-equilibrium adverse-pressure-gradient turbulent boundary layers},
	volume = {8},
	issn = {2469-990X},
	url = {https://link.aps.org/doi/10.1103/PhysRevFluids.8.L022602},
	doi = {10.1103/PhysRevFluids.8.L022602},
	language = {en},
	number = {2},
	urldate = {2023-04-11},
	journal = {Physical Review Fluids},
	author = {Pozuelo, Ramón and Li, Qiang and Schlatter, Philipp and Vinuesa, Ricardo},
	month = feb,
	year = {2023},
	pages = {L022602},
}

@article{schlatter2010assessment,
  author  = {Schlatter, P. and {\"O}rl{\"u}, R.},
  title   = {Quantifying the interaction between large and small scales in wall-bounded turbulent flows: A note of caution},
  journal = {Physics of Fluids},
  year    = {2010},
  volume  = {22},
  number  = {5},
  pages   = {051704}
}

@article{bandyopadhyay1984coupling,
  author  = {Bandyopadhyay, P. R. and Hussain, A. K. M. F.},
  title   = {The coupling between scales in shear flows},
  journal = {Physics of Fluids},
  year    = {1984},
  volume  = {27},
  number  = {9},
  pages   = {2221--2228}
}

@article{hutchins2007large,
  author  = {Hutchins, N. and Marusic, I.},
  title   = {Large-scale influences in near-wall turbulence},
  journal = {Philosophical Transactions of the Royal Society A},
  year    = {2007},
  volume  = {365},
  number  = {1852},
  pages   = {647--664}
}

@article{ganapathisubramani2012amplitude,
  author  = {Ganapathisubramani, B. and Hutchins, N. and Monty, J. P. and Chung, D. and Marusic, I.},
  title   = {Amplitude and frequency modulation in wall turbulence},
  journal = {Journal of Fluid Mechanics},
  year    = {2012},
  volume  = {712},
  pages   = {61--91}
}

@article{marusic2010predictive,
  author  = {Marusic, I. and Mathis, R. and Hutchins, N.},
  title   = {Predictive model for wall-bounded turbulent flow},
  journal = {Science},
  year    = {2010},
  volume  = {329},
  number  = {5988},
  pages   = {193--196}
}

@article{duvvuri2015triadic,
  author  = {Duvvuri, S. and McKeon, B. J.},
  title   = {Triadic scale interactions in a turbulent boundary layer},
  journal = {Journal of Fluid Mechanics},
  year    = {2015},
  volume  = {767},
  pages   = {R4}
}

@article{chung2010large,
  author  = {Chung, D. and McKeon, B. J.},
  title   = {Large-eddy simulation of large-scale structures in long channel flow},
  journal = {Journal of Fluid Mechanics},
  year    = {2010},
  volume  = {661},
  pages   = {341--364}
}

@article{jacobi2013phase,
  author  = {Jacobi, I. and McKeon, B. J.},
  title   = {Phase relationships between large and small scales in the turbulent boundary layer},
  journal = {Experiments in Fluids},
  year    = {2013},
  volume  = {54},
  number  = {3},
  pages   = {1--13}
}

@article{mathis2011predictive,
  author  = {Mathis, R. and Hutchins, N. and Marusic, I.},
  title   = {A predictive inner--outer model for streamwise turbulence statistics in wall-bounded flows},
  journal = {Journal of Fluid Mechanics},
  year    = {2011},
  volume  = {681},
  pages   = {537--566}
}

@article{mathis2011relationship,
  author  = {Mathis, Romain and Marusic, Ivan and Hutchins, Nicholas and Sreenivasan, K. R.},
  title   = {The relationship between the velocity skewness and the amplitude modulation of the small scale by the large scale in turbulent boundary layers},
  journal = {Physics of Fluids},
  year    = {2011},
  volume  = {23},
  number  = {12},
  pages   = {121702},
  doi     = {10.1063/1.3671738}
}

@article{Vallikivi2015TurbulentNumber,
    title = {{Turbulent boundary layer statistics at very high Reynolds number}},
    year = {2015},
    journal = {Journal of Fluid Mechanics},
    author = {Vallikivi, M. and Hultmark, M. and Smits, A. J.},
    volume = {779},
    doi = {10.1017/jfm.2015.273},
    issn = {14697645}
}

@article{pozuelo2022adverse,
	title = {An adverse-pressure-gradient turbulent boundary layer with nearly constant up to},
	volume = {939},
	issn = {0022-1120, 1469-7645},
	url = {https://www.cambridge.org/core/product/identifier/S002211202200221X/type/journal_article},
	doi = {10.1017/jfm.2022.221},
	language = {en},
	urldate = {2023-04-11},
	journal = {Journal of Fluid Mechanics},
	author = {Pozuelo, Ramón and Li, Qiang and Schlatter, Philipp and Vinuesa, Ricardo},
	month = may,
	year = {2022},
	pages = {A34},
}

@article{vinuesa2017pressure,
  title={Pressure-gradient turbulent boundary layers developing around a wing section},
  author={Vinuesa, Ricardo and Hosseini, Seyed M and Hanifi, Ardeshir and Henningson, Dan S and Schlatter, Philipp},
  journal={Flow, turbulence and combustion},
  volume={99},
  pages={613--641},
  year={2017},
  publisher={Springer}
}

@article{Jaroslawski2023LSB,
  title={Disturbance growth in a laminar separation bubble subjected to free-stream turbulence},
  author={Jaroslawski, Tomek and Forte, Maxime and Vermeersch, Olivier and Moschetta, Jean-Marc and Gowree, Erwin R},
  journal={Journal of Fluid Mechanics},
  volume={956},
  pages={A33},
  year={2023},
  publisher={Cambridge University Press}
}

@article{jaroslawski2023boundary,
  title={Boundary layer forcing on a rotating wing at low Reynolds numbers},
  author={Jaroslawski, Tomek and Forte, Maxime and Moschetta, Jean-Marc and Gowree, Erwin R},
  journal={Experiments in Fluids},
  volume={64},
  number={3},
  pages={58},
  year={2023},
  publisher={Springer}
}

@article{lee2017large,
  title={Large-scale motions in turbulent boundary layers subjected to adverse pressure gradients},
  author={Lee, Jae Hwa},
  journal={Journal of Fluid Mechanics},
  volume={810},
  pages={323--361},
  year={2017},
  publisher={Cambridge University Press}
}

@article{gambuzza2021effects,
  title={The effects of free-stream turbulence on the performance of a model wind turbine},
  author={Gambuzza, Stefano and Ganapathisubramani, Bharathram},
  journal={Journal of Renewable and Sustainable Energy},
  volume={13},
  number={2},
  year={2021},
  publisher={AIP Publishing}
}

@book{volino1997new,
  title={A new model for free-stream turbulence effects on boundary layers},
  author={Volino, Ralph J},
  volume={78705},
  year={1997},
  publisher={American Society of Mechanical Engineers}
}
\newpage

\end{document}